%% file: main.tex
\documentclass[preprint,authoryear]{elsarticle}
\usepackage[utf8]{inputenc}
\usepackage{amsmath,amsfonts,graphicx,tabularx}
\usepackage{multirow}
\usepackage{xcolor}
\usepackage{tabularx}
\usepackage{url}
\usepackage{natbib}
\usepackage{todonotes}
\usepackage{rotating}
\usepackage{appendix}
\usepackage{pdfpages}
\usepackage{adjustbox}
\usepackage{fancyvrb}
\usepackage{listings}
\usepackage{glossaries}

\makeglossaries

\newglossaryentry{support set}
{
    name=Support set,
    description={The small set of data that helps define each new sub-task (here, 5 example sounds, and the background sound between)}
}

\newglossaryentry{query set}
{
    name=Query set,
    description={The data for which predictions are to be generated in each sub-task (here, one or more long audio clips)}
}
\newglossaryentry{FSL}
{
    name=FSL,
    description={Few-shot learning}
}
\newglossaryentry{SED}
{
    name=SED,
    description={Sound event detection}
}
\newglossaryentry{Meta-learning}
{
    name=Meta-learning,
    description={Methods that leverage experience on previous tasks to speed-up new learning, by improving the performance of the learner}
}
\newglossaryentry{FSED}
{
    name=FSED,
    description={Few-shot bioacoustic sound event detection }
}
\newglossaryentry{DL}
{
    name=DL,
    description={Deep Learning}
}
\newglossaryentry{ML}
{
    name=ML,
    description={Machine learning}
}
\newglossaryentry{PCEN}
{
    name=PCEN,
    description={Per-channel Energy Normalization}
}

\newcommand{\sidewaystablelabel}[2]{\multirow{#1}{*}{\rotatebox{90}{\textbf{#2}}}}

\date{}

\begin{document}

\begin{frontmatter}

\title{
Learning to detect an animal sound from five examples
}


\journal{journal}

\author[instQMUL,EQFA]{Ines Nolasco} 
\author[instQMUL,EQFA]{Shubhr Singh} 
\author[instQMUL,EQFA]{Veronica Morfi}
\author[instLS2N]{Vincent Lostanlen}
\author[instKON,instMPIAB,instCASCB]{Ariana Strandburg-Peshkin}
\author[instLaSalle]{Ester Vidaña-Vila}
\author[instBiotopia]{Lisa Gill}
\author[instAGH]{Hanna Pamuła}
\author[instSalford]{Helen Whitehead}
\author[instSurrey]{Ivan Kiskin}
\author[instEcoscience,instSyracuse,instWHOI]{Frants H. Jensen}
\author[instOxford]{Joe Morford}
\author[instQMUL]{Michael G. Emmerson}
\author[instQMUL]{Elisabetta Versace}
\author[instMPIAB,instKON,instCASCB]{Emily Grout}
\author[instSurrey]{Haohe Liu}
\author[instNBC,instTIU]{Dan Stowell}

\affiliation[instNBC]{organization={Naturalis Biodiversity Center},
city={Leiden},
country={The Netherlands}}

\affiliation[instTIU]{organization={Tilburg University},
city={Tilburg},
country={The Netherlands}}

\affiliation[instQMUL]{organization={Queen Mary University of London},
city={London},
country={UK}}

\affiliation[instCNRS]{organization={Centre National de la Recherche Scientifique (CNRS)},
city={},
country={France}}

\affiliation[instLS2N]{organization={Nantes Universite, Ecole Centrale Nantes, CNRS},
addressline={LS2N, UMR 6004},
city={F-44000 Nantes},
country={France}}

\affiliation[instKON]{organization={Biology Department, University of Konstanz},
addressline={Universit\unexpanded{ä}tsstrasse 10},
city={Konstanz},
country={Germany}}

\affiliation[instCASCB]{organization={Centre for the Advanced Study of Collective Behaviour, University of Konstanz},
addressline={Universit\unexpanded{ä}tsstrasse 10},
city={Konstanz},
country={Germany}}

\affiliation[instMPIAB]{organization={Department for the Ecology of Animal Societies, Max Planck Institute of Animal Behavior},
addressline={B\unexpanded{ü}cklestrasse 5},
city={Konstanz},
country={Germany}}

\affiliation[instBiotopia]{organization={BIOTOPIA Naturkundemuseum Bayern },
city={Munich},
country={Germany}}

\affiliation[instAGH]{organization={AGH University of Science and Technology },
addressline={al. Adama Mickiewicza 30},
city={Krakow},
country={Poland}}

\affiliation[instLaSalle]{organization={La Salle Campus Barcelona, Universitat Ramon Llull},
addressline={Sant Joan de La Salle, 42,},
city={Barcelona},
country={Spain}}

\affiliation[instSalford]{organization={School of Science, Engineering and Environment,University of Salford},
city={Manchester},
country={UK}}

\affiliation[instSurrey]{organization={Centre for Vision, Speech and Signal Processing, FEPS, University of Surrey},
city={Surrey},
country={UK}}

\affiliation[instOxford]{organization={The Oxford Navigation group, Dept. of Zoology, University of Oxford, },
city={Oxford},
country={UK}}

\affiliation[instEcoscience]{organization={Aarhus University, Department of Ecoscience},
addressline={Frederiksborgvej 399},
city={4000 Roskilde},
country={Denmark}}

\affiliation[instSyracuse]{organization={Syracuse University},
addressline={107 College Place},
city={Syracuse, NY 13244},
country={USA}}

\affiliation[instWHOI]{organization={Woods Hole Oceanographic Institution},
addressline={266 Woods Hole Rd},
city={Woods Hole, MA 02543},
country={USA}}


\affiliation[EQFA]{Equal first authors.}

\begin{abstract}
Automatic detection and classification of animal sounds has many applications in biodiversity monitoring and animal behaviour.
In the past twenty years, the volume of digitised wildlife sound available has massively increased, and automatic classification through deep learning now shows strong results.
However, bioacoustics is not a single task but a vast range of small-scale tasks (such as individual ID, call type, emotional indication) with wide variety in data characteristics, and most bioacoustic tasks do not come with strongly-labelled training data.
The standard paradigm of supervised learning, focussed on a single large-scale dataset and/or a generic pre-trained algorithm, is insufficient.
In this work we recast bioacoustic sound event detection within the AI framework of \textit{few-shot learning}. We adapt this framework to sound event detection, such that a system can be given the annotated start/end times of as few as 5 events, and can then detect events in long-duration audio---even when the sound category was \textit{not known} at the time of algorithm training.
We introduce a collection of open datasets designed to strongly test a system's ability to perform few-shot sound event detections, and we present the results of a public contest to address the task.
We show that prototypical networks are a strong-performing method, when enhanced with adaptations for general characteristics of animal sounds.
We demonstrate that widely-varying sound event durations are an important factor in performance, as well as non-stationarity, i.e.\ gradual changes in conditions throughout the duration of a recording.
For fine-grained bioacoustic recognition tasks without massive annotated training data, our results demonstrate that few-shot sound event detection is a powerful new method, strongly outperforming traditional signal-processing detection methods in the fully automated scenario.
\end{abstract}


\end{frontmatter}
\printglossaries
\section{Introduction}

Machine listening, defined as the application of machine learning to audio content analysis, has untapped potential in the life sciences, applied to animal vocalisations.
Because animal vocalisations vary systematically across species, across social/environmental/emotional contexts, and across individuals \citep{Marler:2004,Brown:2017}, machine listening has the potential to provide crucial information on animal populations and communities as well as on individuals and their behavioral states.
Hence, automated detection and analysis of animal vocalisations is not only valuable for our understanding of sound production but also for diverse research fields including animal behaviour, animal welfare, neuroscience and ecology  \citep{Gillings:2021,Riede:2018,Caiger:2020,Gillespie:2009}.
Recent advances in consumer electronics have considerably lowered the cost and weight of digital audio acquisition, thus allowing deployment of autonomous recording units at large spatiotemporal scales \citep{hill2018audiomoth,roe2021australian,sethi2020safe}.
However, massively distributed bioacoustic surveys have resulted in a ``data deluge'', where data collection outgrows information management.
This issue is not limited to scientific research, where audio corpora serve to conduct statistical hypothesis testing.
Difficulties in handling, analysing and interpreting large amounts of data also extend to applied fields in which animals can be monitored using sound: farming, conservation, and wind energy, to name a few.

Since the beginning of the 21st century, the need for large-scale analyses of animal sounds has spurred the emergence of ``computational bioacoustics'' approaches, complementary to human surveys \citep{Stowell:2018}.
Methods have often been inspired by developments in neighbouring subfields of machine listening---music information retrieval and speech technology---as well as by computer vision.
In this regard, the breakthrough of deep learning in automatic speech recognition, around the year 2012, has profoundly influenced the orientation of computational bioacoustics research \citep{hinton2012deep}.
In particular, most deep learning systems for bioacoustics are trained as sound event \textit{classifiers}: given a short audio excerpt, usually of constant duration, they return an element within a predefined class.
This approach is derived from the ``phone classification task'' used in speech analysis, with animal vocalisations in lieu of human utterances, and a species-specific catalogue in lieu of a phonetic alphabet \citep{Ganchev:2017}. 

However, the paradigm of supervised sound event classification based on speech is reaching its limits in computational bioacoustics.
Indeed, the extrapolation between speech to other animal sounds is difficult and limited, due to
differences in sound duration and units of interest, differences in context and taxonomy, as well as differences in recording conditions, among others.
First, detecting the start and end time of animal sounds has a key role in community ecology, since so much of the structure lies in call-and-response and other patterns of influence \citep{Stowell:2016,Logue:2016}.
Secondly, bioacoustic practitioners operate at many different levels of granularity, from coarse (e.g.,\ species classification) to fine (e.g.,\ distinguishing call types or syllables from one individual); whereas speech science relies on limited levels of granularity where human phonemes or words  are the fundamental units.
Thirdly, non-human animal sounds are acquired with a plethora of diverse equipment, including far-field, on-body, and underwater, whereas speech sounds are typically acquired with an individual device, that is usually controlled by the person speaking. 

A main limitation in bioacoustics is the lack of a unified framework that can be applied to different vocalisations. Today, the literature on computational bioacoustics is fragmented into subdomains: marine versus terrestrial, individual versus species identification, handheld versus fixed equipment, and so forth \citep{Frazao:2020,Kahl:2020,Linhart:2022}. 
Overall, all these subdomains share a common definition of what constitutes a ``sound event'': i.e., a recognisable auditory perception with an onset and offset. 
However, the spectrotemporal characteristics underlying these events vary dramatically across species and domains.
Thus, bioacoustic event detection does not appear as a single ``big-data'' problem; but instead, as a juxtaposition of many small-data problems, each currently addressed by specialised systems.
The field benefits from the common coarse-scale task of species classification, which has provided a clear and useful focus to drive computational bioacoustics into the deep learning era \citep{Joly:2019,Kahl:2021}.  Yet, systems trained for coarse-scale tasks, even with massive data, do not automatically acquire the ability to make fine-grained or local distinctions, and must be further trained or customised \citep{Lauha:2022,vanHorn:2021}. Thus, much recent work (re)trains deep learning systems anew for each specific new task.

Such fragmentation hinders the practical usability of deep learning in bioacoustics, and thus in the life sciences at large.
Indeed to date, the success of deep neural networks in the supervised regime depends on the availability of a massive corpus of audio examples for the sound events of interest, paired with human annotations.
Yet, temporally-precise and fine-grained annotation of audio demands expertise, and is thus costly and time-consuming.
In many cases, the obstacle is not only to acquire annotations, but also the audio examples themselves: e.g. for rare species, remote locations, or costly equipment.
Furthermore, these numerous small-data scenarios remain outside the scope of digital bioacoustic archives, such as Xeno-Canto and the Macaulay Library. 


In this article, we aim to develop a unified method that works across the many subdomains of computational bioacoustic sound event detection (SED).
The benefit of doing so resides in the development of a robust and versatile system that could serve the scientific community at large.
Hence, we assembled a collection of 14 small-scale datasets, between 10 minutes and 10 hours in duration.
Each of them reflects a genuine but slightly different application setting and are obtained from completely different sources.
The main originality of our work is that, instead of training 14 individual machine listening systems (one per dataset), we train a single system to detect sound events on many different datasets, in which each dataset has a different category of sound event to be detected---that category only defined at ``query time''.
Furthermore, when being evaluated on an audio file, the system is prompted with the first five occurrences of the sound event of interest.
This paradigm of machine learning is known as ``few-shot learning''(\gls{FSL}) \citep{Snell:2017,Wang:2020review}.

Stated otherwise, our hypothesis is that bioacoustic event detectors can take advantage of whichever bioacoustic datasets are available at training time, and then generalise from a few (five) examples of the new target at deployment time.
This is difficult under a standard supervised paradigm because the training set may not reflect real-world deployment conditions, nor cover all sound categories of possible interest.
For these reasons, we place the concept of domain adaptation at the heart of the few-shot learning paradigm in bioacoustics: our goal is not only to learn a detector from limited labeled data but also to learn domain-agnostic representations of animal sounds which can readily adapt to unforeseen recording conditions (cf.\ \citet{beery2018recognition} in computer vision).

In order to diversify methods and accelerate progress, we have organised an open-science challenge for a community of researchers named DCASE: ``Detection and Classification of Acoustic Scenes and Events''. 
The challenge was open to everyone and consisted of public datasets, evaluation metrics, documentation, and baseline systems.

In this paper we formulate bioacoustic sound event detection (SED) as a few-shot learning task. We describe \gls{ML} systems customised to this new task (published openly as baseline methods), and we report on a public data challenge conducted over two years to generate and evaluate novel algorithmic solutions.
\footnote{Development data: \url{https://zenodo.org/record/6482837} \\Evaluation data \url{https://zenodo.org/record/6517414} \\Code: \url{https://github.com/c4dm/dcase-few-shot-bioacoustic/}}
We evaluate various dimensions of the \gls{ML} paradigms that have been put forward for this task, and explore their ability to adapt to aspects of bioacoustic data presented in our datasets.
Our study demonstrates that few-shot SED is a feasible way forward in bioacoustics.

\subsection{Related work}

Few-shot \textit{classification} has been investigated generally, and also for audio (acoustic) data \citep{Snell:2017,Pons:2019,Shi:2020,Naranjo:2022}. However, \gls{SED} has different requirements from classification: typically, the desired output includes the onset and offset times for each detected event \citep{Mesaros:2016}, roughly similar to the ``object detection'' task in computer vision.

One important insight behind few-shot learning is that of \textit{meta-learning} (``learning to learn''), or the idea of leveraging past experience to speed-up new learning by improving the performance of the learner \citep{schaul2010metalearning}. One approach to meta-learning is training a system across many loosely-similar tasks/datasets, such that the system is then well-configured to generalise from a few examples of a novel class \citep{Ravi:2017,Wang:2020review}. This depends on a system learning something of the implicit commonalities and analogies across the tasks, which might then influence an algorithm's learnt feature extraction, or its measure of similarity between data points, for example.
Related work in computer vision explores the challenge of fine-grained classification and object detection in images from camera traps in novel conditions \citep{beery2018recognition}.

\citet{vanHorn:2021} introduced a wildlife image dataset with multiple subsets each defining a different binary question.
This has a similar meta-learning spirit as our work, with the aim that a sophisticated first stage of ``representation learning'' across multiple tasks can make future tasks simple.
However, unlike the few-shot setting that we use, a lightweight (shallow) classifier must be trained for each new question from a non-trivial number of positive/negative examples.

A previous data-driven challenge on animal vocalisation audio detection was focused on birds \citep{Stowell:2019}. That challenge also aimed to generalise robustly to conditions not seen in the training data but was simpler than ours, in that it did not require systems to predict event onset or offset times, only presence/absence; it stayed within the framework of supervised classification rather than generalising from examples of new categories; and it didn't include as broad a range of animal taxa.

\section{Method}

\subsection{Task formulation}

\begin{figure}[t]
    \centering
    \includegraphics[width=0.6\textwidth]{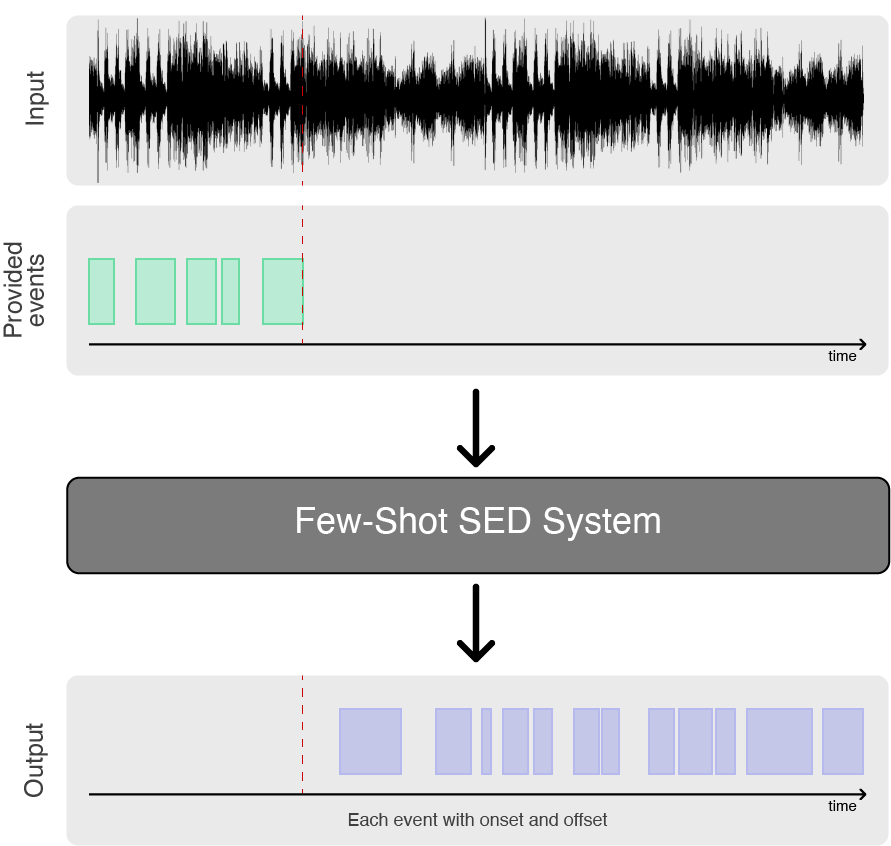}\hfill
    \includegraphics[width=0.3\textwidth]{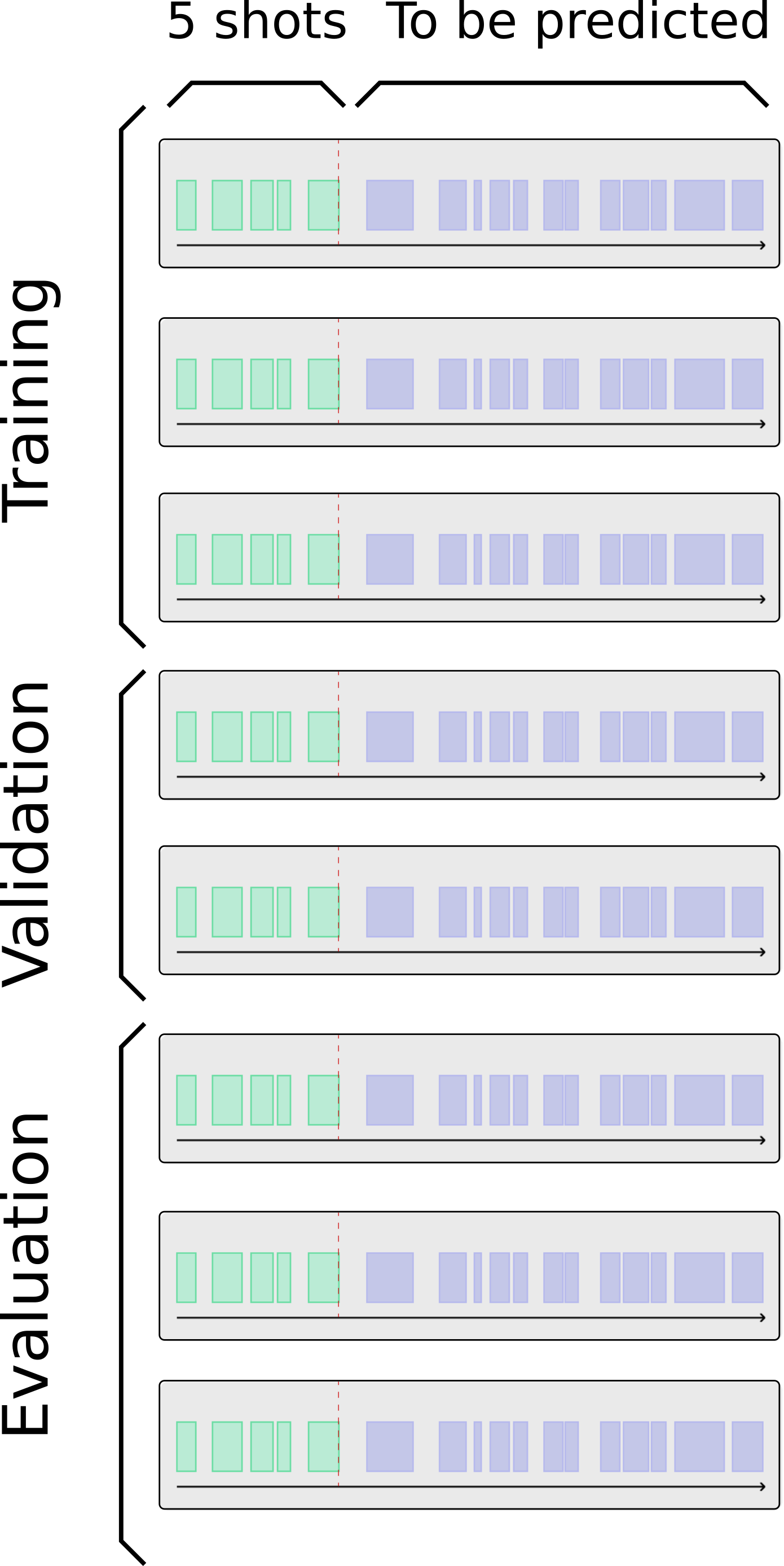}
    \caption{(a) Few-shot sound event detection: the first 5 sound events are given as examples---in standard supervised learning they would be considered the training set---and the remainder must then be detected. (b) Few-shot sound event detection as a \textit{meta-learning} problem. Each of our datasets represents a different but related few-shot task. The overall goal is to use the training and validation datasets collectively to train or otherwise develop a system that, when presented with 5 sound events from any of the evaluation datasets, can perform well at detecting the remaining events.}
    \label{fig:fsSED}
    \label{fig:fewshotmetalearning}
\end{figure}

We formulate few-shot bioacoustic sound event detection (\gls{FSED}) as follows:

\textit{Given one long audio recording (or multiple audio recordings), as well as annotations on the onset and offset time for each of the \textit{first five} sound events of interest, identify the onset and offset times for all other such sound events in the recording(s)} (Figure \ref{fig:fsSED}a).

To train a system for this using meta-learning, we make use of multiple bioacoustic datasets  (Figure \ref{fig:fewshotmetalearning}b) representing a range of taxa and recording conditions, each annotated with a different target sound category (see next section).

Note that we do not consider multiple classes in one dataset \citep{Naranjo:2022,Mesaros:2019}: each dataset represents a single-class problem. Other sounds are undoubtedly present in almost all audio recordings, but these are considered to be background noise (clutter/distractor events). Our formulation is easily extended to multiple classes in a scene by applying inference separately for each category of interest.

We choose few-shot rather than one-shot learning because animal sounds of interest often cover a range of variability: for example, there may be multiple call types in the set of sounds of interest, or calls from multiple distinct individuals within a group. Five as the number of examples is a conventional choice in few-shot learning, but could vary \citep{Snell:2017,Ravi:2017,Pons:2019}.

Note also that we choose to use the \textit{first} occurring events as examples, rather than a randomly-selected set. This reflects typical practice in bioacoustics, in that acoustic data are typically labeled in contiguous time segments which may or may not be fully representative of the entire data set, and should be tractable for future users of few-shot acoustic systems.
It offers one benefit, that an algorithm may make use of the strong assumption that the periods between the first five examples fall into the negative class.
It also aligns with common scenarios, such as manually labelling data during a pilot phase and then deploying a recognition system to automatically label incoming data.
However, it also presents a risk: the first sequence of examples may be similar to each other in some way which is not representative of the whole set of events, for example if the acoustic environment or animal behaviours exhibit non stationary characteristics but change over time.

\subsection{Datasets}
\label{sec:datasets}

A conventional machine learning experiment uses a single dataset partitioned into three subsets, used for training, validation, and evaluation (test).
In the few-shot formulation, we also divide the data into these three partitions (training/validation/evaluation), but each in fact contains \textit{multiple datasets}, and each dataset represents one example of a few-shot task.
Within each dataset, there are one or more audio files, each accompanied by a CSV text file giving the start time, end time and label of the targeted audio events.
The label can be POS for a positive example of the target class, NEG for a negative example (background or non-target sound event), or UNK for unknown cases, where the human annotator(s) were unsure whether a sound event should be considered in the positive class.
Such UNK cases may often occur in complex wildlife sound scenes; our chosen strategy was to explicitly label these cases, allowing algorithm designers to make their own decisions on how to handle them, but to exclude UNK time-regions from the evaluation measures (described later) since their correct label is ambiguous.
For each dataset, the first five POS events are the ``few shots'' from which the rest should be inferred.

A \textit{development set} was provided for the task when the challenge was launched, consisting of the predefined training and validation sets to be used for system development. The development set consists of datasets from multiple sources with audio recordings and associated reference annotations in our specified format. More specifically, for the training set multi-class temporal annotations were provided for each recording (with multiple POS/NEG/UNK columns in the data, one per class), while for the validation set single-class temporal annotations (POS/UNK) were provided for each recording. 

A separate \textit{evaluation set} was kept for evaluating the performance of the systems. During the task, only the five POS event annotations were provided for each of the recordings for the class of interest. The developed systems had to use those five annotated events and then learn to detect the same type of events throughout the rest of the recording. The true annotations for the rest of the recording were kept private for evaluation.

Table \ref{tab:datasets} presents an overview of all the datasets in the development and evaluation sets used in the 2022 edition of the challenge.

\begin{table}[thp]
\centering
\begin{adjustbox}{center}
%
%
    \begin{tabular}{r|c|c|c|c|c|c|c|c|c}
          &  &  &  &  \# & Total & \# & \# & &  Mean event  \\
         \textbf{2022}& Dataset & Taxon & Mic type &  Files & duration & Labels & Events & Density & length (sec)\\
         \hline
        \sidewaystablelabel{4}{Training}
        & BV & Birds & fixed & 5 & 10 hours & 11 & 9026& 0.038& 0.15\\
        & HT & Mammals & on-body & 5& 5 hours & 5 & 611&0.047&1.42\\
        & MT & Mammals & on-body & 2 & 70 mins & 4 & 1294& 0.042 & 0.15\\
        & JD & Birds & on-body & 1 & 10 mins & 1 & 357 & 0.062& 0.11\\
        & WMW & Birds & various & 161 & 5 hours  & 26 & 2941 &0.25& 1.54\\
        \hline
        \sidewaystablelabel{2}{Val.}
        & HB & Insects & handheld & 10 & 2.4 hours & 1 & 712 & 0.67 &11.67\\
        & PB & Birds & fixed &  6 & 3 hours & 2 & 292 & 0.003 & 0.11\\
        & ME & Mammals & handheld & 2 & 20 mins & 2 & 73& 0.011 &0.19\\
        \hline
        \sidewaystablelabel{6}{Evaluation}
        &  CHE & Birds & fixed & 18 & 3 hours & 3 & 2550 & 0.263 & 1.94\\
        & DC & Birds & fixed &  10 & 95 mins & 3 & 967 & 0.350 & 1.66\\
        & CT & Mammals & on-body & 3  & 48 mins  & 3 & 365 & 0.017 & 0.16\\
        & MS & Birds & fixed &  4 & 40 mins & 1 & 1087& 0.084 & 0.18\\
        & QU & Mammals (marine) & on-body &  8 & 74 mins & 1 & 3441 & 0.045 & 0.06\\
        & MGE & Birds & fixed &  3 & 32 mins & 2 & 1195 &0.194 & 0.27\\
    \end{tabular}
\end{adjustbox}
\caption{Information on each dataset used in the 2022 challenge. ``Density'' is calculated as in signal processing: (total duration of events) / (total duration of audio), thus values close to 0 are sparse, and close to 1 are dense.}
    \label{tab:datasets}
\end{table}

\textbf{BirdVox-DCASE-10h (BV):}
The BV dataset contains five audio files from four families of  (birds). The recordings were obtained through four different autonomous recording units, each lasting two hours.
These autonomous recording units are all located in Tompkins County, NY, US.
They follow the same hardware specification: the Recording and Observing Bird Identification Node (ROBIN) developed by the Cornell Lab of Ornithology \citep{Lostanlen:2018}.
All recordings were acquired in 2015, during the fall migration season.
An expert ornithologist, Andrew Farnsworth, has annotated flight calls from four families of passerines, namely: American sparrows, cardinals, thrushes, and New World warblers.
The annotator found 2,662 flight calls from 11 different species in total.
These flight calls have a duration in the range 50--150 milliseconds and a fundamental frequency in the range 2--10 kHz.

\textbf{Hyenas (HT):} 
The HT dataset contains five recordings from hyenas. Spotted hyena vocalisation data were recorded on custom-developed audio tags (DTAG) designed by Mark Johnson and integrated into combined GPS/acoustic collars (Followit Sweden AB) by Frants Jensen and Mark Johnson, \cite{1190131}. Collars were deployed on female hyenas of the Talek West hyena clan at the MSU-Mara Hyena Project (directed by Kay Holekamp) in the Masai Mara, Kenya as part of a multi-species study on communication and collective behavior.
Spotted hyenas are a highly social species that live in ``fission-fusion" groups where group members range alone or in smaller subgroups that split and merge over time. Hyenas use a variety of types of vocalisations to coordinate with one another over both short and long distances \citep{Lehmann:2020}.
Recordings used as part of this task contain a variety of different vocalisations which were identified and classified into types based on the established hyena vocal repertoire \citep{leblond2021group}. 
There is no overlap between the vocalisations annotated in the two sets. Fieldwork was carried out from November 2016 - February 2017 by Kay Holekamp, Andrew Gersick, Frants Jensen, Ariana Strandburg-Peshkin, Benson Pion, Morgan Lucot, and Rebecca LeFleur; labelling was done by Kenna Lehmann and colleagues.

\textbf{Meerkats (MT, ME):}
The MT and ME datasets contains two recordings each from meerkats. Recordings used in this task were acquired at the Kalahari Meerkat Project (Kuruman River Reserve, South Africa; directed by Marta Manser and Tim Clutton-Brock), as part of a multi-species study on communication and collective behavior. Recordings of the development set (MT) were recorded on small audio devices (TS Market, Edic Mini Tiny+ A77, 8 kHz) integrated into combined GPS/audio collars which were deployed on multiple members of meerkat groups to monitor their movements and vocalisations. Recordings of the evaluation set (ME) were recorded by an observer following a focal meerkat with a Sennheiser ME66 directional microphone (44.1 kHz) from a distance of typically less than 1 m.
Meerkats are a highly social mongoose species that live in stable social groups and use a variety of distinct vocalisations to communicate and coordinate with one another. Recordings were carried out during daytime hours while meerkats were primarily foraging and include several different call types. The meerkat vocal repertoire has been well characterised based on previous research, allowing calls to be reliably classified by human labellers \citep{Manser:1998, Manser:2014}.   Fieldwork was carried out by Ariana Strandburg-Peshkin, Baptiste Averly, Vlad Demartsev, Gabriella Gall, Rebecca Schaefer and Marta Manser; and the recordings were labelled by Baptiste Averly, Vlad Demartsev, Ariana Strandburg-Peshkin, and colleagues.

\textbf{Jackdaws (JD):}
The JD dataset contains a 10-minute on-bird sound recording (22.05 KHz) of one male jackdaw during the breeding season in 2015. In a multi-year field study (Max-Planck-Institute for Ornithology, Seewiesen, Germany), wild jackdaws were equipped with small backpacks containing miniature voice recorders (Edic Mini Tiny A31, TS-Market Ltd., Russia) to investigate the vocal behaviour of individuals interacting with their group and behaving freely in their natural environment.
Jackdaws are corvid songbirds that usually breed, forage and sleep in large groups, but form a pair bond with the same partner for life. 
Fieldwork was conducted by Lisa Gill, Magdalena Pelayo van Buuren and Magdalena Maier. Sound files were annotated by Lisa Gill, based on a previously established video-validation in a captive setting \citep{Stowell:2017}.

\textbf{Western Mediterranean Wetlands Bird Dataset (WMW):} The WMW dataset contains 161 files with bird sounds from 20 endemic species that are typically inhabitants of the ``Aiguamolls de l'Empordà" natural park in Girona, Spain. The audio files that compose this dataset were originally retrieved from the Xeno-Canto portal  \citep{Vellinga:2015}. Xeno-Canto is a portal in which citizens can upload wildlife sounds. As the audio files are collected by a wide community of people, the recording devices used for gathering data can be different in every audio file.  
Depending on the species, audios contain vocalisations such as bird calls or songs; or sounds such as bill clapping (\textit{Ciconia ciconia} species) or drumming (\textit{Dendrocopos minor} species). 
For the WMW dataset, Juan Gómez-Gómez, Ester Vidaña-Vila and Xavier Sevillano  manually cleaned and labelled downloaded audio files using the Audacity software \citep{Gomez-Gomez:2023}. 

\textbf{HumBug (HB):} The HB dataset contains sounds of lab-cultured \emph{Culex quinquefasciatus} mosquitoes from Oxford, UK, and various species captured in the wild in Thailand, placed into plastic cups \citep{Li:2018}. Mosquitoes produce sound both as a by-product of their flight and as a means for communication and mating. Fundamental frequencies vary in the range of 150 to 750 Hz \citep{Kiskin:2020}. As part of the HumBug project, acoustic data was recorded with a high specification field microphone (Telinga EM-23) coupled with an Olympus LS-14. The recordings used in this challenge are a subset of the datasets marked as \emph{`OxZoology'} and \emph{`Thailand'} from HumBugDB \citep{Kiskin:2021}\footnote{ \url{https://github.com/HumBug-Mosquito/HumBugDB/}}. 

\textbf{Polish Baltic Sea bird flight calls (PB):}
The PB dataset consists of six 30-minute recordings of bird flight calls recorded along the Polish Baltic Sea coast.  Three autonomous recording units were used with the same hardware settings (Song Meters SM2, Wildlife Acoustics, Inc). They were deployed close to each other ($<$100m) - near the lake, on the dune, and on the forest clearing - to provide diverse acoustic background. The recordings were acquired during the 2016, 2017 and 2018 fall migration seasons. The recordings are the excerpt from Hanna Pamuła's project, focused on the acoustic monitoring of birds migrating at night along the Polish Baltic Sea coast \citep{Pamula:2022a,Pamula:2022b}. The passerines night flight calls were annotated by Hanna Pamuła. 
In each recording, only one bird species is the target class: song thrush, \emph{Turdus philomelos} (3 recordings); blackbird, \emph{Turdus merula} (3 recordings). The usual fundamental frequency range for calls of the chosen species is 5--9 kHz.

\textbf{Transfer-Exposure-Effects dataset (CHE):}
The CHE dataset contains bird vocalizations from  the Chornobyl Exclusion Zone (CEZ). Data were collected using unattended acoustic recorders (Songmeter 3) to capture the Chornobyl soundscape and investigate the longterm effects of the nuclear power plant accident on the local ecology. This dataset comes from the Transfer Exposure-Effects (TREE) research project\footnote{\url{https://tree.ceh.ac.uk/}}. 
To date, the study has captured approximately 10,000 hours of audio from the CEZ.
The fieldwork was designed and undertaken by Mike Wood (University of Salford), Nick Beresford (UK Centre for Ecology \& Hydrology) and Sergey Gashchak (Chornobyl Center). Common Chiffchaff (\textit{Phylloscopus collybita}) and Common Cuckoo (\textit{Cuculus canorus}) vocalisations were manually annotated and labelled from these recordings by Helen Whitehead.

\textbf{BIOTOPIA Dawn Chorus (DC):}
The DC dataset used as part of the evaluation set stems from dawn chorus recordings, made using Zoom H2 recorders at 44.1 KHz, at three different locations in Southern Germany (Haspelmoor, Munich’s Nymphenburg Schlosspark, and Nantesbuch).
Many bird species produce vocalisations during the entire day, but their vocally most active period by far usually occurs around dawn. This natural phenomenon of \textit{dawn chorus} has received a lot of attention in biological studies, and also appears to be the perfect time window for species detection, as it provides the largest likelihood of most individuals of the same and of different species signalling. Yet the sheer complexity of undirected dawn chorus recordings have made automatic species classification extremely difficult, leaving this potentially rich source of acoustically determined species data largely untapped. 
The Dawn Chorus project is a worldwide citizen science and arts project bringing together amateurs and experts to experience and record the dawn chorus at their doorstep, to draw a global picture of bird biodiversity through sound. Recordings were obtained by by Moritz Hertel and Rudi Schleich. The vocalisations of three target species (Common cuckoo, \emph{Cuculus canorus}; European robin, \emph{Erithacus rubecula}; Eurasian wren, \emph{Troglodytes troglodytes}) were annotated by Lisa Gill.

\textbf{Coati (CT):} The CT dataset contains audio recordings collected from two adult females from the same group on Barro Colorado Island, Panama in March 2020. These individuals wore 
collars which recorded high-resolution GPS data with an external attachment of a small audio recording device (TS Market, Edic Mini Tiny+ A77, 22.05 KHz). Audio data were recorded during their active foraging period in daytime hours when a variety of social and aggressive calls are commonly emitted.
Coatis are omnivorous diurnal mammals that live in stable social groups consisting of females and related juvenile and subadult males. Coatis produce a number of call types that are used across a variety of different behavioural contexts. The documentation of their complete vocal repertoire is currently under development. The target calls used in this dataset are growls, chitters and chirp-grunts. Several other call types that might be confused with the targets were captured in the recordings which present the main challenging aspect of this data. Fieldwork was carried out by Emily Grout, Josué Ortega and Ben Hirsch. Calls were labelled by Emily Grout using Adobe Audition.

\textbf{Manx Shearwater (MS):} The MS dataset contains vocalizations from Manx Shearwater individuals, which are procellariform  seabirds that breed in dense island colonies in the North Atlantic, mostly in the British Isles, and winter in the South Atlantic off the South American coast. In a multi-year study, Audiomoth recorders were placed in burrows on Skomer Island to record the vocalisations of both adult Manx shearwaters and chicks during the breeding season. Adult Manx shearwaters make loud, distinctive vocalisations while present at their breeding colony in various contexts: in duets with their partner in their nesting burrow, to broadcast from their burrow, and during flight. Pairs of Manx shearwaters raise single chicks in underground burrows, regularly visiting the breeding colony at night to feed their chick. During these visits, the chick vocalises to 'beg' for food from the parent shearwater; these vocalisations typically comprise bouts of short high-pitched 'peeps'. Fieldwork was undertaken by various members of the Oxford Navigation Group (OxNav), associated with the Oxford University Department of Biology and led by Professor Tim Guilford. Annotation of individual chick begging vocalisations was carried out by Joe Morford using Sonic Visualiser (Release 4.0.1; Queen Mary, University of London); these vocalisations, therefore, represent the target class in this dataset.

\textbf{Dolphin Quacks (QU):}
The QU dataset contains recordings from Bottlenose dolphin sounds in their natural habitat obtained using sound-and-movement recording DTAGs \cite{johnson2003digital}, attached with suction cups to bottlenose dolphins by Frants Jensen in collaboration with Drs. Peter Tyack, Vincent Janik, Laela Sayigh, Randall Wells and the Sarasota Dolphin Research Program. All tags were deployed during routine health assessments conducted by the Sarasota dolphin research project and under a National Marine Fisheries Service research permit to Dr. Randall Wells of Chicago Zoological Society.
Bottlenose dolphins are highly acoustic animals with an expansive repertoire of acoustic signals used for social interactions. 
Male bottlenose dolphins (\textit{Tursiops truncatus}) in Sarasota form close pair bonds with other males that help them consort with females during the mating season. 
The target class is Quacks, which are short, low-frequency narrowband signals (around 100 ms duration and main energy below a few kHz) \cite{simard2011low}, 
and emitted at relatively high rates by one or both males in the alliance, often with 100s of quacks in a single short vocal bout. 
Individual quacks were labelled by Austin Dziki and validated by Frants Jensen.

\textbf{Chick calls (MGE):}
The MGE dataset contains three 10-minute recordings from three 1-day old domestic chicks (\textit{Gallus gallus}). Vocalisations have been recorded and annotated in the Prepared Minds Lab (Queen Mary University of London\footnote{\url{https://www.preparedmindslab.org/home}}) by Dr Versace's staff (Shuge Wang, Michael Emmerson, Laura Freeland, Elisabetta Versace). Individual chicks have been recorded in the controlled environment of the laboratory, a 24-48 hours after hatching. Chickens are a precocial social bird species and upon hatching they establish a strong attachment to their social companions, via a process called imprinting, where acoustic information strengthens affiliative responses \cite{versace2017spontaneous}. 
During and after the imprinting process, chicks vocalise signaling that they are in close proximity to their social partners (i.e. pleasure calls) or that they are distant or separated from them (i.e. contact calls).
The data gathered in the dataset present uneven time distribution. Calls typically have a short duration (100-400 milliseconds).
In the dataset, only pleasure calls were annotated in recordings from chicks one and two, only contact calls were annotated in recordings from chicks three.

\begin{figure*}[th]
\includegraphics[width=\textwidth,clip,trim=20mm 30mm 9mm 35mm]{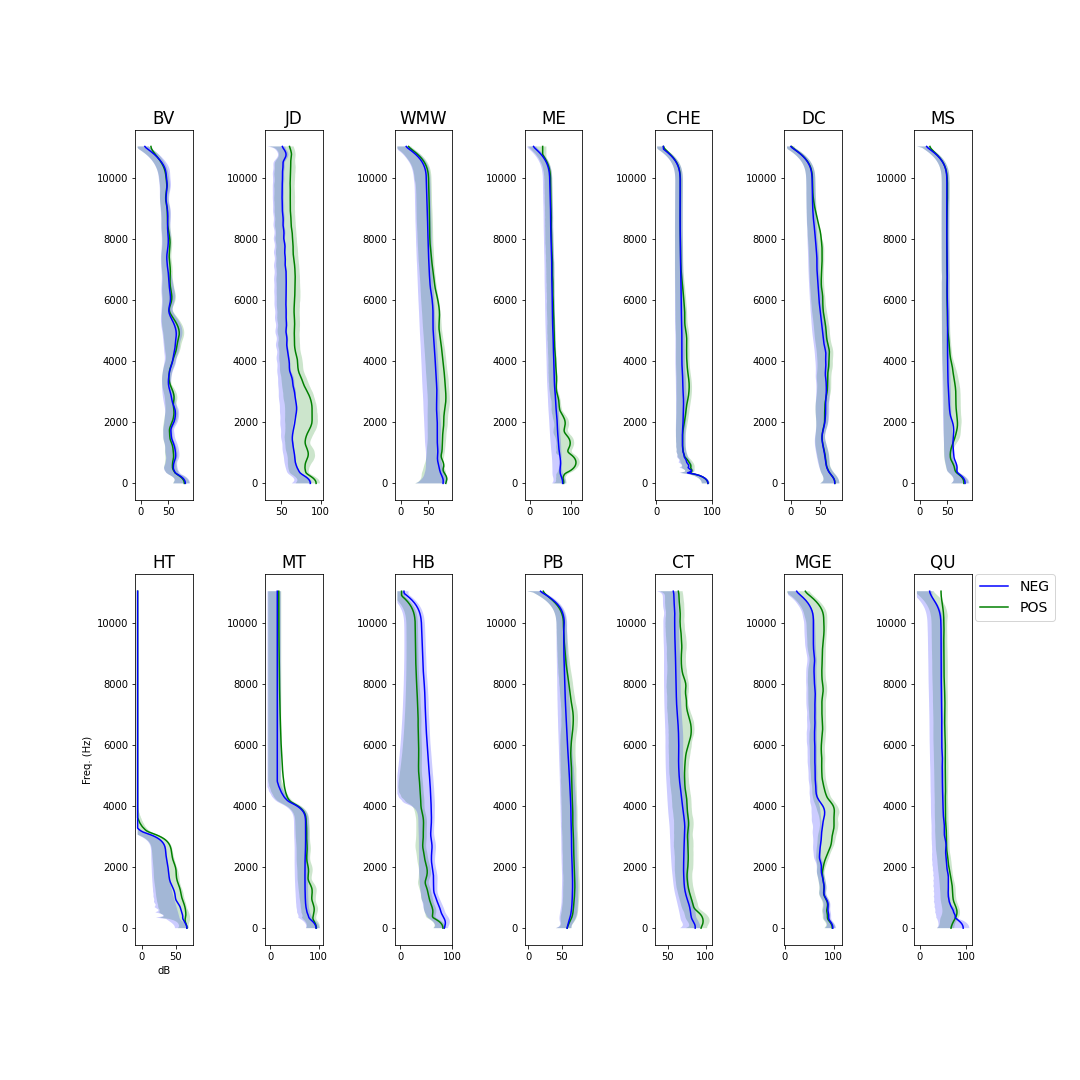}
\caption{Spectral summary profiles of each dataset. For each frequency, we show mean and 90\% confidence intervals of the energy distribution, for the foreground (POS events) and negative regions (background and non target sounds) separately.}
\label{fig:specprofiles}
\end{figure*}

\begin{figure*}[th]
\includegraphics[width=\textwidth,clip,trim=3.9cm 6cm 0cm 6.35cm]{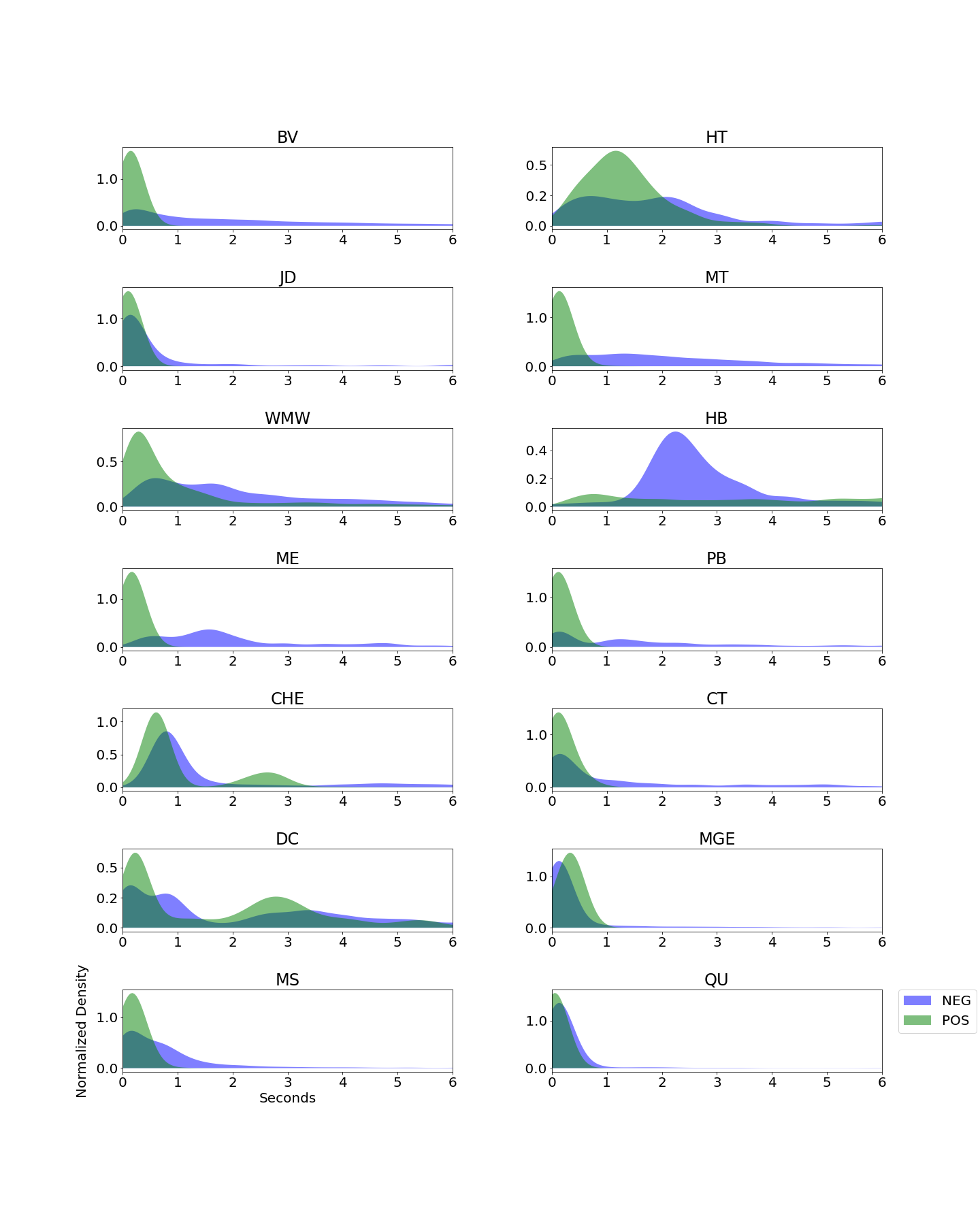}
\caption{Temporal profiles of each dataset. We show the empirical distributions (kde smoothed) of durations of marked regions, for the foreground (POS events) and negative regions (all non-POS regions) separately.}
\label{fig:timeprofiles}
\end{figure*}

\begin{figure*}[th]
\includegraphics[width=\textwidth]{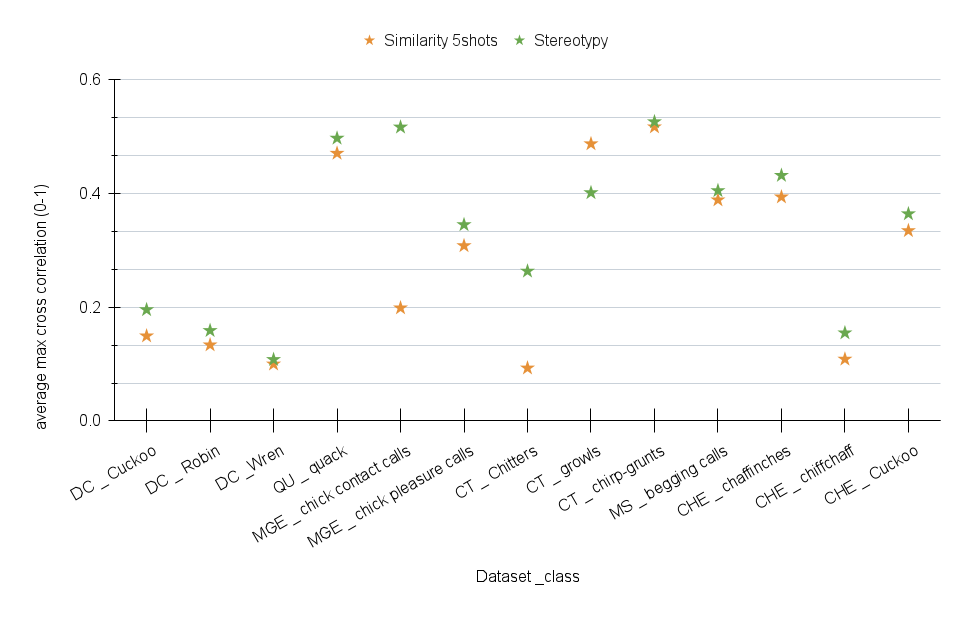}
\caption{Values of similarity between the annotated calls and the first 5 events (shots), and stereotypy for each class in the evaluation set. Classes are indicated in the horizontal axis by DatasetName\_ClassName.  
The similarity metric is based on the average maximum cross correlation between events. It ranges between 0 and 1, where values closer to 1 represent higher similarity. (details in \ref{app:sterotypy}).}
\label{fig:similaritywith5shots}
\end{figure*}

To summarise, these datasets together represent some of the wide variety of bioacoustic SED tasks, and were selected to give broad coverage of some of the key axes of variation, such as rate of occurrence of the target sound, length of calls,  background noise (SNR), taxa, etc. (Table \ref{tab:datasets}).
Descriptive analysis of the datasets further illustrates the variation in temporal and spectral characteristics, for the target sounds as well as the background soundscapes (Figure \ref{tab:spectrograms}, Figure \ref{fig:specprofiles} and Figure \ref{fig:timeprofiles}).
The datasets represent diverse challenges for the few-shot SED systems that are trained and evaluated on them.
For each dataset, the provided 5 events are used to specify the class of target sounds.
The extent to which a small set of calls can be representative depends on various factors including stereotypy - the degree of how stereotyped are the calls, and vocabulary size. 
To approximately quantify this, for each class in the evaluation set, we calculate similarity between sound events. We do this between the selected five events and the remaining events, as well as for the annotated calls more generally (Figure \ref{fig:similaritywith5shots}).
Together with the SNR and the sparsity/density of call events, this stereotypy aspect is expected to be one of the axes of variation among our datasets.
(details in \ref{app:sterotypy})

\input{spectrograms/table_spectrograms.tex}

\clearpage
\subsection{Baseline methods}
\label{sec:baseline}

We propose two systems as baselines, representative of standard good-quality methods that can be applied to the task, and against which to measure the performance of novel submitted methods.
One is an approach commonly used in bioacoustics based on spectrogram cross-correlation and the other is a deep learning approach based on prototypical networks, which have been used in other FSL work. 

\subsubsection{Template matching (cross-correlation)}
\label{ssec:tempbase}

Signal-processing methods have been used for decades to detect events of possible interest in audio data \citep{Towsey:2012,Gillespie:2009}.
Common approaches include energy thresholding, which can work in low-noise scenarios only,
and template matching, usually based on cross-correlation (matched filtering) of waveforms or spectrograms. Template matching can work well in noisy audio, providing the target signal is acoustically (a) distinct from the background sounds and (b) stereotyped, i.e.\ not strongly varying in character.
We thus expect template matching to work well in some of the scenarios we study, but to perform very poorly in others.

Our baseline cross-correlation method is based on scikit-image’s \texttt{match\_template} function applied to spectrograms: it uses fast, normalised cross-correlation to find instances of a template in an image, returning values ranged between -1.0 and 1.0, with higher values corresponding to higher correlation. Our few-shot template matching method computes cross-correlation across the time axis between each of the events (shots) provided for a file and the rest of the recording. A different detection threshold is set for each audio file based on the max value of the cross-correlation results between the shots provided. Peak picking is performed on the results of the template matching algorithm, with any peak above the threshold corresponding to the center of a detected event in that recording. Borders of the predicted event are assumed to align with the beginning and end of the template when it matches. Each of the 5 templates is used separately for matching, and the resulting event predictions are collapsed into a single binary prediction vector which will produce the final events predicted for the class of interest.

\subsubsection{Prototypical networks}

\label{ssec:protobase}
Our second baseline is based on prototypical networks, a deep learning technique whose training procedure is designed especially for few-shot learning \citep{Snell:2017}.
The networks are trained using \textit{episodic training}: each ``episode'' is configured as an ``$N$-way-$k$-shot'' classification task, where $N$ denotes the number of classes and $k$ the number of known samples per class. In the present work $k=5$, and $N=2$ when there is only one sound event type to consider, in which case the two classes then represent active and inactive.
Prototypical networks have previously been evaluated as highly promising for few-shot audio classification tasks \citep{Pons:2019}.

A \textit{prototype} in this method is a coordinate in some vector representation, which is calculated as a simple centroid (mean) of the coordinates for each of the $k$ examples.
The training data consist of a \textit{\gls{support set}} $S$ consisting of $k$ labelled samples from each class, with the remaining samples comprising the \textit{\gls{query set}} $Q$.
Prototypical networks compute a class prototype $c_n$ through an embedding function $f_\phi :\mathbb{R}^D\xrightarrow{}\mathbb{R}^M$ with learnable parameters $\phi$. In our baseline system $D=128$ and $M=64$, and $f_\phi$ is a neural network.
The prototype for class $n$ is computed as the mean of the embedded support points belonging to that class:
\begin{equation}
    c_n = \frac{1}{k}\sum_{(x_i) \in S_n} f_\phi(x_i)
\end{equation}
where $S_n$ represents the subset of $S$ from class $n$.

Then, for each sample $x_q$ from the query set, a distance function is used to calculate the Euclidean distance of $x_q$ from each prototype, following which a softmax function over the distances produces a distribution over the classes.
This directly implies that training the neural network to optimise these distances should move prototypes and their corresponding query points closer together in the embedding space created by $f_\phi$, and further away from non-matching points.
In other words, the training procedure creates a general representation in which similar sounds are close to each other.
Nearest-neighbour algorithms such as k-means can then be used to label future data points---even those from novel categories, after a simple procedure of calculating the prototype of a novel category as the centroid of its $k$ shots.

During evaluation, we adopt a binary classification strategy inspired by~\citet{Wang:2020}. The first 5 positive (POS) annotations are used for calculation of positive class prototype and the rest of the audio file is treated as the negative class, based on the assumption that the positive class is relatively sparse in the recording. We randomly sample time regions from the negative class to calculate the negative prototype. 
Each query sample is predicted to have the target sound active, if its embedding coordinate is closer to the positive prototype than the negative.
The prediction process for each file is repeated 5 times, with the negative prototype created by random sampling each time. The final prediction probability for each query frame is the average of predictions across all iterations. Finally, post-processing is applied to the outputs in order to remove possible false positives. For each audio file, predicted events with shorter duration than 60\% of the duration of the shortest shot provided for that file are removed.

\subsection{Novel approaches}



Deep learning models for few-shot learning problems can be broadly categorized into two approaches: meta-learning and transfer learning. Meta-learning or \textit{learning to learn}~\citep{thrun2012learning} focuses on learning priors from previous experiences in order to efficiently adapt to new tasks. 

Prototypical networks, introduced above, is a well-known example of such meta-learning. These, as well as matching networks~\citep{vinyals2016matching}, have performed well in few-shot learning tasks across both image and audio domain.

As mentioned earlier, meta-learning based methods rely on the assumption that the tasks belong to a single distribution, for example metric learning based methods require the tasks all coming from a similar domain such that there exists a uniform metric that could work across tasks~\citep{wang2019hybrid}. However, in real world scenarios this assumption does not always hold such as in case of our task where the datasets vary in terms of species, recording conditions and microphones, essentially rendering the problem as a cross-domain few-shot learning. In such cases, a hybrid meta-learning approach towards the task may be required, which moves beyond the assumption that future tasks are well-represented by the set of training tasks. A few hybrid methods are as follows:
\begin{itemize}
    \item \textbf{Cross-domain few-shot learning } - Very few methods specifically designed to account for cross-domain scenarios have been previously explored.  Feature-wise transformation layers were introduced in \citet{tseng2020cross} for augmenting the features  using affine transforms, in order to adapt to domain shift across tasks. In \citet{dong2019domain},  an adversarial network based model is used for one-shot domain adaption from source to target domain. 
    \item \textbf{Transductive few-shot learning} - Meta learning methods aim to learn on scarce data in order to generalise to unseen tasks, which makes the problem fundamentally difficult. In order to mitigate the difficulty, transductive based methods utilise \textit{the information present in the unlabeled examples from the query set} to adapt the model and improve its predictions. In \citet{liu2018learning} , the samples in support and query set are jointly modelled as nodes of a graph and the prediction on query set is conducted by label-propagation algorithm.  In \citet{hou2019cross}, a cross-attention based map is learnt between support set and query set in order to make predictions on individual query examples. 

\end{itemize}

Alternatively, transfer learning based methods rely on adapting to a new task through the transfer of knowledge from a related task that has already been learned ~\citep{parnami2022learning}. First, a deep learning model is  trained on large training set of base class and then fine-tuned on a few examples of the novel class.  Fine-tuning on a few examples of the novel class can often lead to poor generalisation, hence  techniques have to be adopted in order to avoid overfitting. For example,  in \citet{wang2021few}, a dynamic few-shot learning approach is adopted where an auxiliary model is used as a few-shot classification ``weight generator'' which uses an attention map between the existing classification weight vector of the base classes and the few-shot examples of the novel classes. SimpleShot~\citep{wang2019simpleshot}  uses a pretrained deep network to get feature embeddings for the input and query set and performs L2 normalisation on the obtained features, subsequently, an Euclidean distance based nearest neighbour classification is performed. A similar approach with cosine-distance was proposed in \citet{chen2020new}.

Through the outcomes of the public challenge, we evaluate some combinations of these novel approaches for the particular domain of bioacoustic SED.

\subsection{Evaluation and public challenge}

\label{sec:eval}
For the evaluation of this task, we employ an event-based F-measure with macro-averaged metric, to evaluate the match between true and predicted events.
The main complexity is related to the detection of a match between ground truth events and predicted events. Traditional approaches use onset detection based metrics and fixed-size evaluation windows \citep{Mesaros:2019}. Given the great variation between datasets and characteristics of the events we want to detect in this task, these approaches are not suitable.
Instead, we use the Intersection over Union (IoU), with 30\% minimum overlap to produce a list of possible matches of the predictions.  
Applied to temporal events we get a list of predicted events that overlap at least 30\% with the ground truth events and thus are candidate matches. 
For each ground truth event, a single best match is selected by applying the Hopcroft-Karp-Karzanov algorithm for bipartite graph matching, a similar procedure as used in the \texttt{sed eval} toolbox.\footnote{\url{http://tut-arg.github.io/sed_eval/generated/sed_eval.util.event_matching.bipartite_match.html}}

In a SED task we can define True Positives (TP) as predicted events that match ground truth events, False Positives (FP) as predicted events that do not match any ground truth events, and False Negatives (FN) as ground truth events that are not predicted. In this task, ground truth events consist of POS events of the class and UNK events that have some uncertainty associated to the assigned class. 
The procedure we employ is:
\begin{enumerate}
\vspace{-0.1cm}
    \item Apply IoU and bipartite graph matching between predicted events and ground truth POS events only, resulting in TP.
    \vspace{-0.1cm}
    \item Apply IoU and bipartite graph matching between remaining predicted events, that did not match with any POS event, and ground truth UNK events only.
    \vspace{-0.1cm}
    \item Compute FP as the number of predicted events that were not matched to either POS or UNK events.
    \vspace{-0.1cm}
    \item Compute FN as the number of POS ground truth events that were not matched by any predicted event.
    \vspace{-0.1cm}
\end{enumerate}
This is applied to each dataset in the evaluation set where we compute the F-score metric. The reported results are the harmonic mean over all the datasets, which is appropriate for combining percentage results, and ensures that a system should perform well across all datasets to achieve a strong score. 

We thus use an averaged F-score as our main summary statistic for each submitted system. To explore system performance in more detail, we also inspect the F-scores per dataset, and per class in each dataset, in particular to examine whether differences in acoustic characteristics correlate with differences in performance.

The F-score metric is designed to summarise how well a system's outputs correspond to the desired outputs. However, there are many factors that affect the usefulness of such outputs, meaning that it is difficult to estimate a technology readiness level from only numerical scores. Hence, in addition to our quantitative analysis, we conduct a qualitative user-oriented analysis of selected system outputs, gathering feedback from expert users (annotators of the datasets).

\section{Results}


\begin{table*}[t]
    \centering
    \small
    \begin{adjustbox}{center}

    \begin{tabular}{l|l|l}
    \textbf{Team} & \multicolumn{1}{l|}{\begin{tabular}[c]{@{}l@{}}\textbf{Evaluation} \\(\textbf{95\% CI)}\end{tabular}} 
    & \multicolumn{1}{l}{\begin{tabular}[c]{@{}l@{}} \textbf{Validation} \end{tabular}}\\
    \hline
    Du\_NERCSLIP \citep{Tang:2022}  
    & \textbf{60.22}  (59.66-60.70)  & 74.4    \\

    Liu\_Surrey \citep{Liu:2022a} 
    & 48.52  (48.18-48.85)  & 50.03  \\

    Martinsson\_RISE  \citep{Martinsson:2022}
    & 47.97  (47.48-48.40)   & 60   \\

    Hertkorn\_ZF \citep{Hertkorn:2022} 
    & 44.98  (44.44-45.42)  & 61.76  \\

    Liu\_BIT-SRCB  \citep{Liu:2022}   & 44.26  (43.85-44.62)  & 64.77   \\

    Wu\_SHNU  \citep{Wu:2022} & 40.93 (40.48-41.30)   & 53.88   \\

    Zgorzynski\_SRPOL \citep{Zgorzynski:2022}  & 33.24  (32.69-33.69)  &  57.2 \\

    Mariajohn\_DSPC \citep{Mariajohn:2022} & 25.66 (25.40-25.91)  & 43.89  \\

    Willbo\_RISE  \citep{Willbo:2022}  & 21.67  (21.32-21.97) & 47.94 \\

    Zou\_PKU  \citep{Yang:2022} & 19.20 (18.88-19.51)  & 51.99 \\

    Huang\_SCUT \citep{Huang:2022} & 18.29 (18.01-18.56)  & 54.63  \\

    Tan\_WHU \citep{Tan:2022} & 17.22  (16.82-17.55)   & 54.53  \\

    Li\_QMUL  \citep{Li:2022} &  15.49 (15.16-15.77) & 47.88  \\
        \hline

    baseline-TempMatch \citep{Morfi:2021} & 12.35 (11.52-12.75)   & 3.37 \\

    baseline-ProtoNet \citep{Morfi:2021} & 5.3 (5.1-5.2)   &  28.45   \\
        \hline

    Zhang\_CQU \citep{Zhang:2022} & 4.34 (3.74-4.56)  & 44.17  \\

    Kang\_ET  \citep{Kang:2022} & 2.82 (2.76-2.87) & -  \\

    \end{tabular}
            \end{adjustbox}
    \caption{2022 $F$-score results (in \%) per team (best scoring system) on evaluation and validation sets.
    Systems are ordered by higher scoring rank on the evaluation set. These results and technical reports for the submitted systems can be found on  task 5 results page \citep{Task5ResultsPage}. An expanded version of this table is available in Supplementary Table \ref{tab:teams2022full}.}
    \label{tab:teams2022}

\end{table*} 

\begin{figure}
    \centering
    \includegraphics[width=\linewidth, trim = 3.2cm 3cm 1cm 0cm, clip]{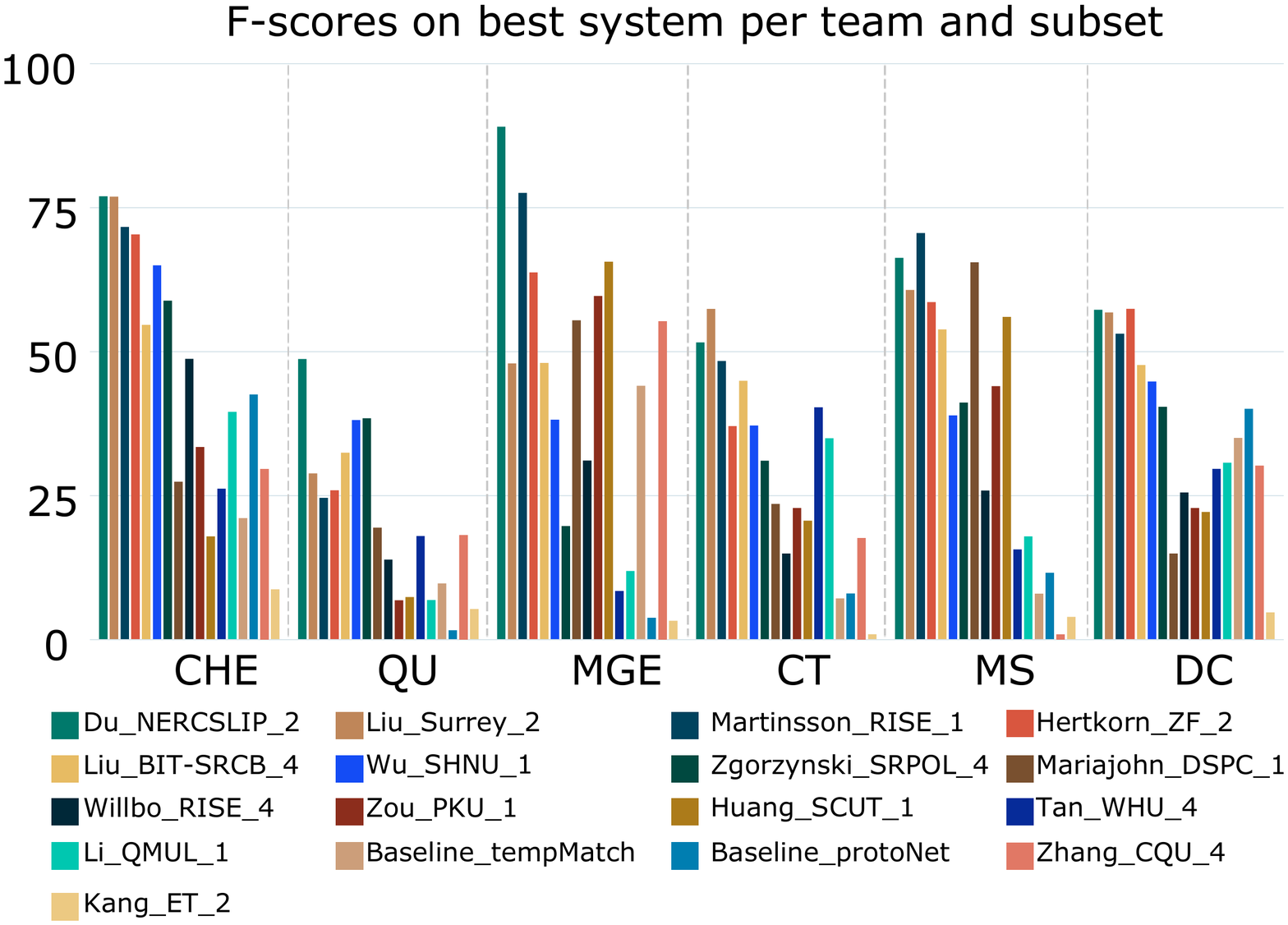}
    \caption{2022 $F$-Score results by dataset. Systems are ordered by overall highest scoring rank on the evaluation set.}
    \label{fig:Fscore_dataset2022}
\end{figure}



\begin{figure}[ht]
    \centering
    \includegraphics[width=0.6\textwidth]{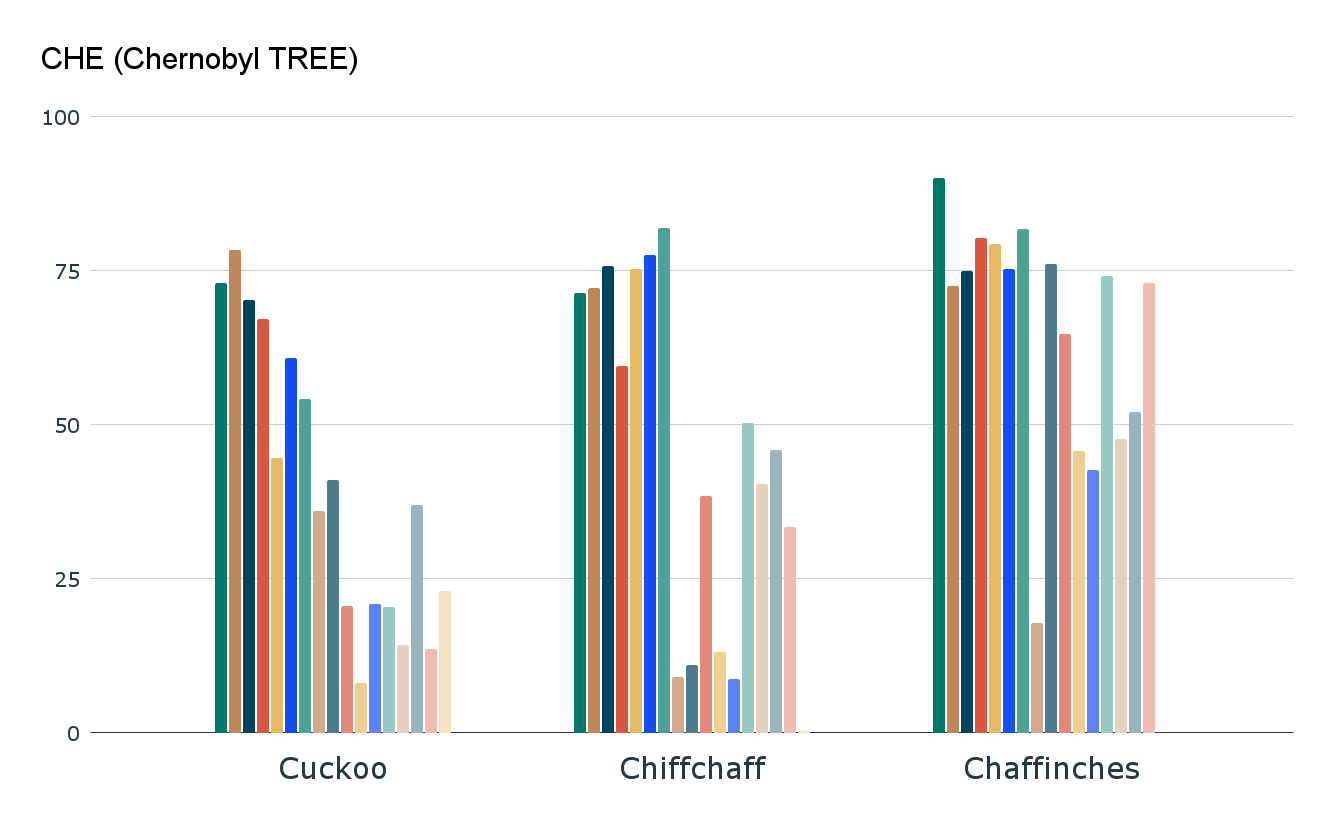}\\
    \includegraphics[width=0.6\textwidth]{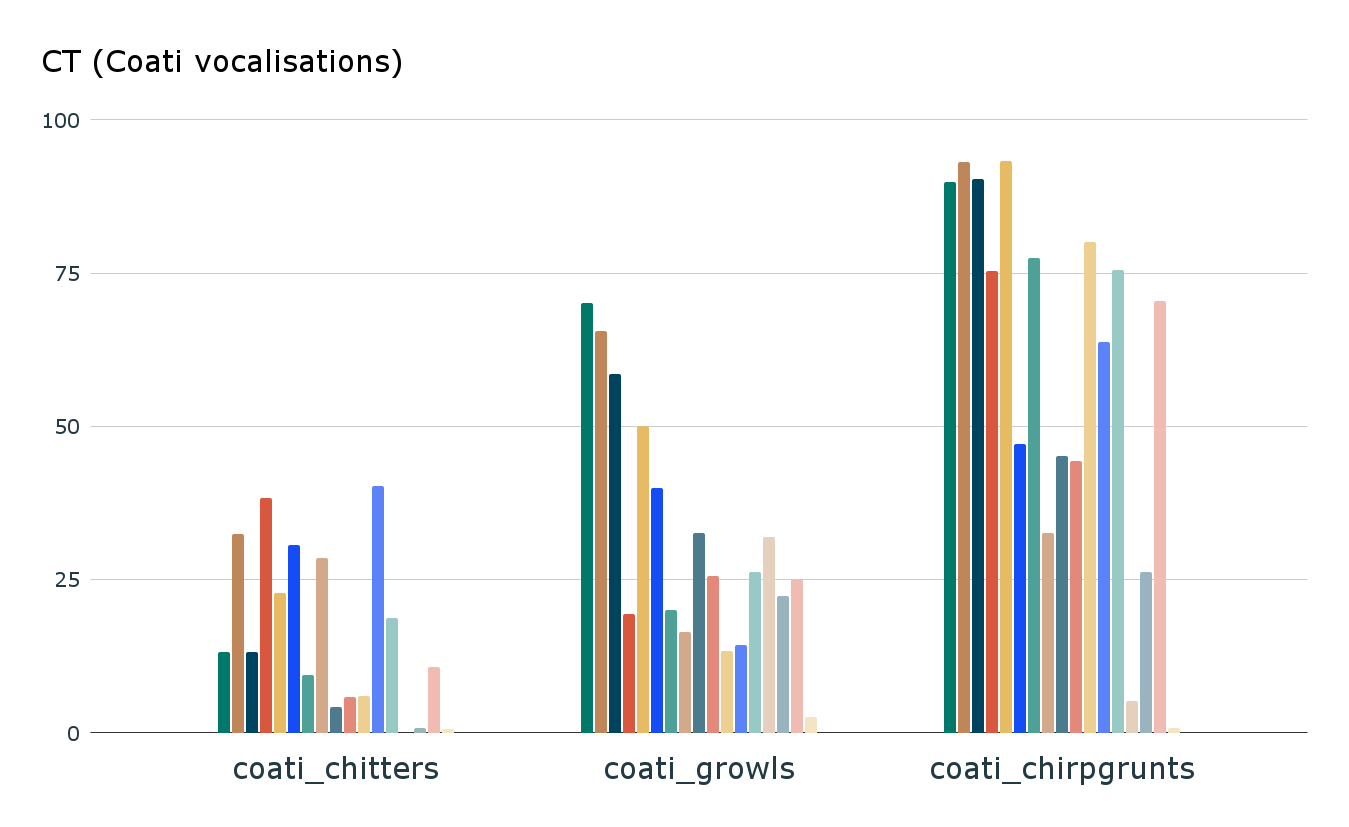} \\
    \includegraphics[width=0.6\textwidth]{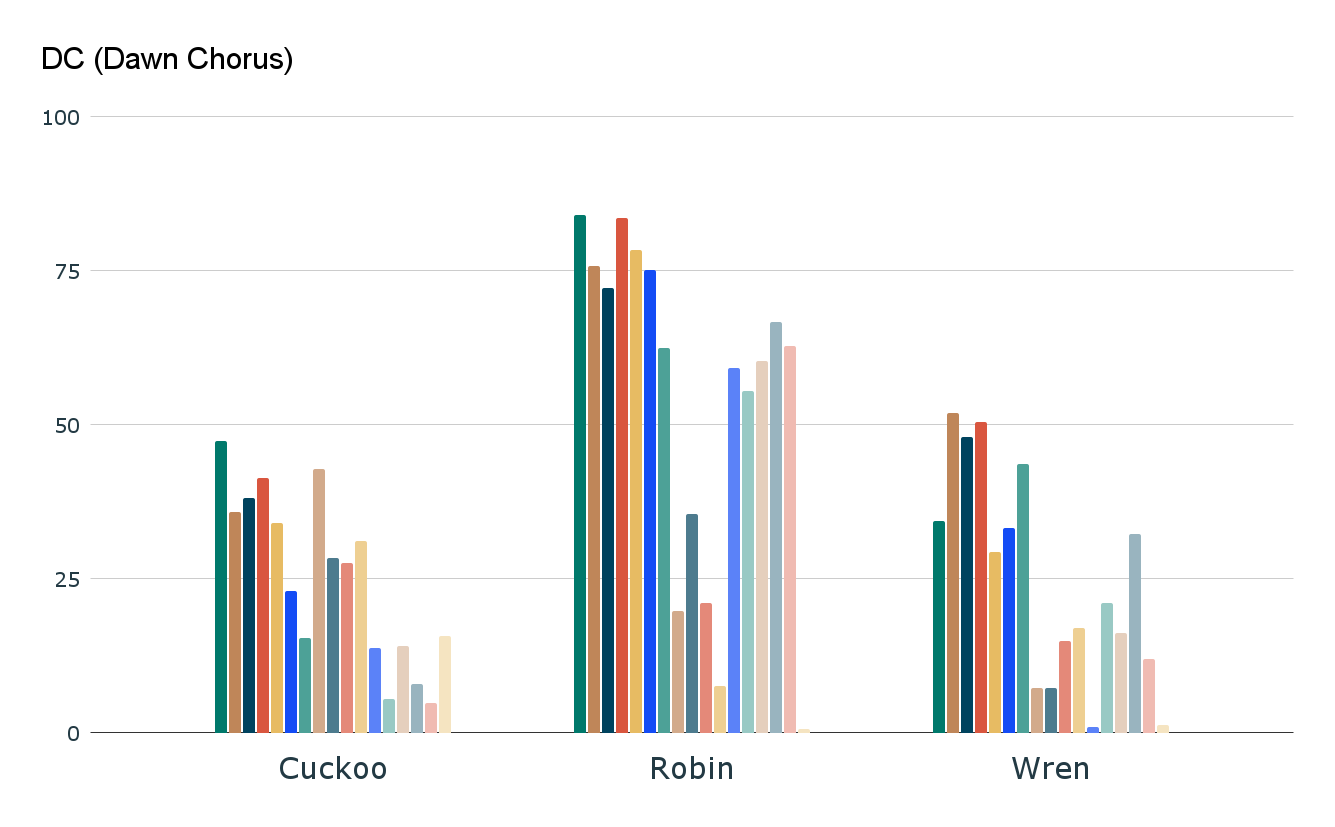}\\
    \includegraphics[width=0.6\textwidth]{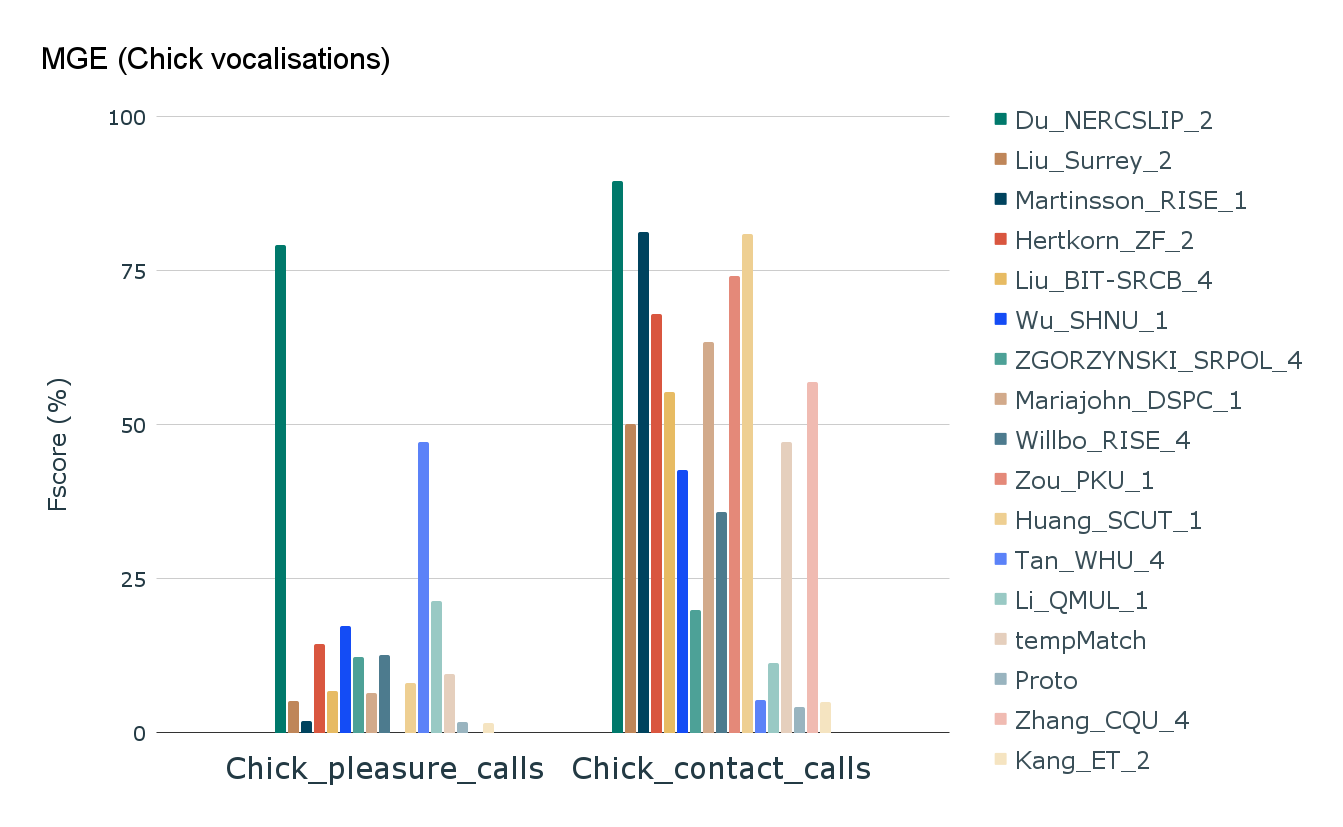} \\
    \caption{F\-score (\%) results by class in the evaluation set. Note that QU and MS datasets only contain a single class and thus are not represented here. The systems are ordered by overall highest scoring rank on the evaluation set}
    \label{fig:fscores_by_class}

\end{figure}

We report here the results from the 2022 edition of our public challenge.
This was in fact the second public edition of the challenge, following a first smaller edition in 2021.
The 2022 results made use of a wider range of evaluation datasets than 2021, and thus a more robust estimation of system performance.
For completeness, a summary of the 2021 outcomes are given in the Supplementary Information.

For the 2022 edition, 15 teams participated submitting a total of 46 systems (Table \ref{tab:teams2022}).
The challenge can be seen to be a difficult one: the baseline systems, and many teams, obtained F-score averages below 25\%.
On the other hand, methods could be designed which reach well over 40\% F-score average, and up to 60\% (Figure \ref{fig:Fscore_dataset2022}).
Such performances were much stronger than expected based on the task difficulty and 2021 results.

The majority of systems adopted a prototypical network approach, perhaps influenced by the baseline code and/or the outcomes of the 2021 edition.
Simple improvements over the baselines were achieved by applying data augmentation techniques and intelligent post-processing.
Better ways to construct the negative prototype were also explored by some teams who reported improved results (\cite{Liu:2022a}, \cite{Martinsson:2022}, \cite{Wu:2022}, \cite{Willbo:2022}).
\textit{Transductive inference}---adapting the learnt feature space at test-time based on the newly-presented positive and negative events---was also applied by some participants (\cite{Liu:2022a}, \cite{Li:2022}, \cite{Tan:2022}, \cite{Yang:2022}).

The highest scoring system implements a frame-level embedding learning approach which confers to the system a high time resolution capability \cite{Tang:2022}. 
The system ranked in second place implements a novel approach designed to optimise the contrast between positive events and negative prototypes \cite{Liu:2022a}. This, together with an adaptive segment length dependent on each target class, works well across all the evaluation sets.

The problem of very different lengths of events across target classes was also directly addressed by other submissions. Both \cite{Martinsson:2022} and \cite{Zgorzynski:2022} implemented an ensemble approach where each individual model focuses on a different input size range.
In \cite{Liu:2022} this is explored through a multi-scale ResNet, and in \cite{Willbo:2022} with a wide ResNet containing many channels.

Finally, it is worth mentioning the system in \cite{Hertkorn:2022}. Their few-shot adaptation was based on fine-tuning alone. The innovation here is related to simple modifications to a CNN-based architecture in order to optimise the use of information, particularly in the frequency axis. Furthermore, by allowing the network to overfit (up to a degree) to the 5 shots, the system achieves surprisingly good performance across all the datasets of the evaluation set.


Inspecting the characteristics of the methods performing most strongly in the challenge, broadly across both editions, we observe some general tendencies (Table \ref{tbl:methodsanalysis}).
Firstly, there is relatively little variation in the acoustic features extracted, and the neural network architecture: most systems use Mel spectrograms with PCEN, and standard CNNs.
However, there is considerable variation in the method of training the network, and performing inference.
There is a roughly equal balance of the two main paradigms: meta-learning with prototypical networks, versus fine-tuning or otherwise adapting a network trained using cross-entropy.

Within both paradigms there are instances of transductive inference \citep{Yang:2021,Liu:2022a}.
The `dynamic few-shot learning' (DFSL) method employed by \citet{Wu:2022} is an alternative approach to query-time adaptation: the feature extraction, and the representation for previously-known classes, is never altered, but at query time the new task is considered to be a new class, whose representation is a weighted sum of those for the previously-known classes. This has the appealing characteristics of combining stability with dynamic adaptation, unlike standard fine-tuning in which care must be taken not to overfit to the new examples.
Despite these innovations, it is notable that multiple teams achieved strong performance without test-time adaptation of the learnt feature space.

Many teams innovated in the way time-regions are selected for training an algorithm, both for computing the positive and negative regions (foreground and background).
Multiple teams made use of \textit{pseudo-labelling} as a way to bootstrap the amount of data presented to the system:
this means using the system to make a first `draft' identification of which regions are positive/negative for the events of interest, and then using that estimated labelling to further train the system \citep{Yang:2021,Tang:2022,Wu:2022}.
Pseudo-labelling has been explored in many machine learning domains for data-poor scenarios.

Successful systems also commonly used explicit methods to control the duration of the detected events. In many cases this consists of postprocessing predictions to delete/merge very short events, or estimating the typical duration from the examples. \citet{Tang:2022} and \citet{Wolters:2021} made use of neural network architectures specifically trained to infer and output region annotations.

Overall, the different approaches submitted illustrate the introduction of ideas to address challenges related to this task: how to deal with very different event lengths; how to construct a negative class when no explicit labels are given for this; and how to bridge the gap between classification and detection for few-shot sound event detection.
These challenges derive from the combination of few-shot learning with sound event detection, and hence are not addressed in standard few-shot learning \citep{Wang:2020review}.

\begin{sidewaystable}
\centering
\begin{tabular}{p{0.5cm}|p{2cm}|p{1.4cm}|p{1.2cm}|p{1.2cm}|p{1.5cm}|p{1.5cm}|p{1.5cm}|p{1.5cm}|p{1.4cm}|p{1.5cm}|p{2cm}}
& & Spectrogr. features & Neural net arch. & Training objective & New class addition & Inference & Feature space changes? & Negatives selection & Positives selection & Segment length technique & Post-processing \\
\hline
\hline
\sidewaystablelabel{2}{Baselines} & Prototypical & Mel +PCEN & CNN & Proto & Proto & Dist:Proto & No & Whole audio & 5 & Delete very short & -- \\
& Template matching & Lin & n/a & n/a & New templates & Cross-correl & No & n/a & 5 + aug & Template length & -- \\
\hline
\sidewaystablelabel{3}{2021 challenge} & Yang 2021 TI & Mel & CNN & x-ent & Retrain (new pos+neg) & Posterior & TI x-ent & Pseudo-neg & Pseudo-pos & & {\raggedright Peak picking, thresholding} \\
& Tang 2021 (SHNU) & Lin +PCEN & CNN & Proto & Proto & Dist:Proto (Attention-weighted) & No & Whole audio & 5 & & Peak picking, median filtering \\
& Anderson 2021 TCD & Mel +PCEN & CNN & Proto & Proto & Dist:Proto & No & Whole audio & 5 & Minimum event length & Probability averaging, median filtering \\
\hline
\sidewaystablelabel{2}{2022 challenge} & Tang 2022 finetune (NERCSLIP) & Mel +PCEN & CNN framewise & x-ent & Finetune last layer & Posterior & Finetune x-ent & Between-the-5 & Pseudo-pos & Adaptive length fixed shift & CRNN event filter \\
& Liu 2022 surrey & Mel +PCEN \& delta-MFCC & CNN & Proto (modifed) & Proto & Dist:Proto & TI, Retrain & Between-the-5 + Pseudo-neg (SpecSim) & 5 & Derived from shots for each class & Split-merge-filter; delete very long/short \\
& Wu 2022 (+Wu 2023 ICASSP) DFSL & Mel +PCEN & CNN (ResNet) & x-ent & DFSL attentive & DFSL attentive & No & Pseudo-neg & Pseudo-pos & & \\
\hline
\sidewaystablelabel{1}{Other} & Wolters 2021 arxiv Perceiver & Mel & CNN +CRNN +Perceiver & Proto +RPN (R-CRNN) & Proto & Dist:Proto & No & n/a & 5 & Region proposal network & \\
\end{tabular}
\caption{Methodological features of various systems of interest. %
Terms: Proto = prototypical network; x-ent = cross-entropy; Dist = distance-based; TI = transductive inference; DFSL = dynamic few shot learning; 5 = the original 5 shots are used; between-the-5 = the spaces between the 5 shots are used; pseudo-pos/pseudo-neg: pseudo-labelling is used to select additional examples. %
}
\label{tbl:methodsanalysis}
\end{sidewaystable}

\subsection{Analysis of dataset dependencies}
The submitted systems exhibit variations in their performance across our datasets (Figure \ref{fig:Fscore_dataset2022}).
The same is true even within datasets at the level of the target class (Figure \ref{fig:fscores_by_class}).
The easiest classes to be detected are CHE\_chaffinches, CT\_chirpgrunts and DC\_robins, where several systems reach above 75\% F-score. On the other side,  CT\_Chitters, DC\_Cuckoo and the QU\_Quacks seem to be the classes where systems struggled the most to make correct predictions.
The disparity in score between systems is also evident. The performance on MGE\_Chick\_Pleasure\_calls is a good example where the DU\_NERCSLIP system shows a significant advantage over the others.

\begin{figure}[ht]
\centering
{\includegraphics[width=0.4\textwidth]{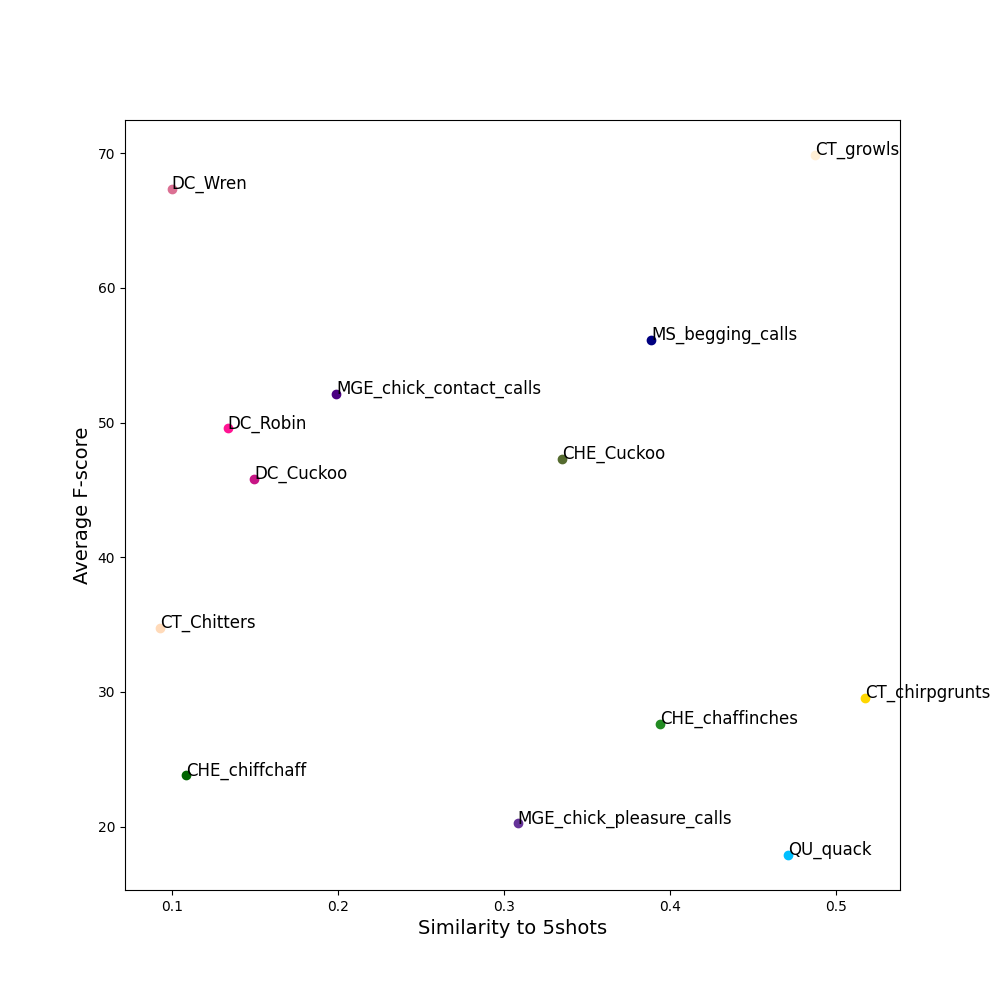}}
{\includegraphics[width=0.4\textwidth]{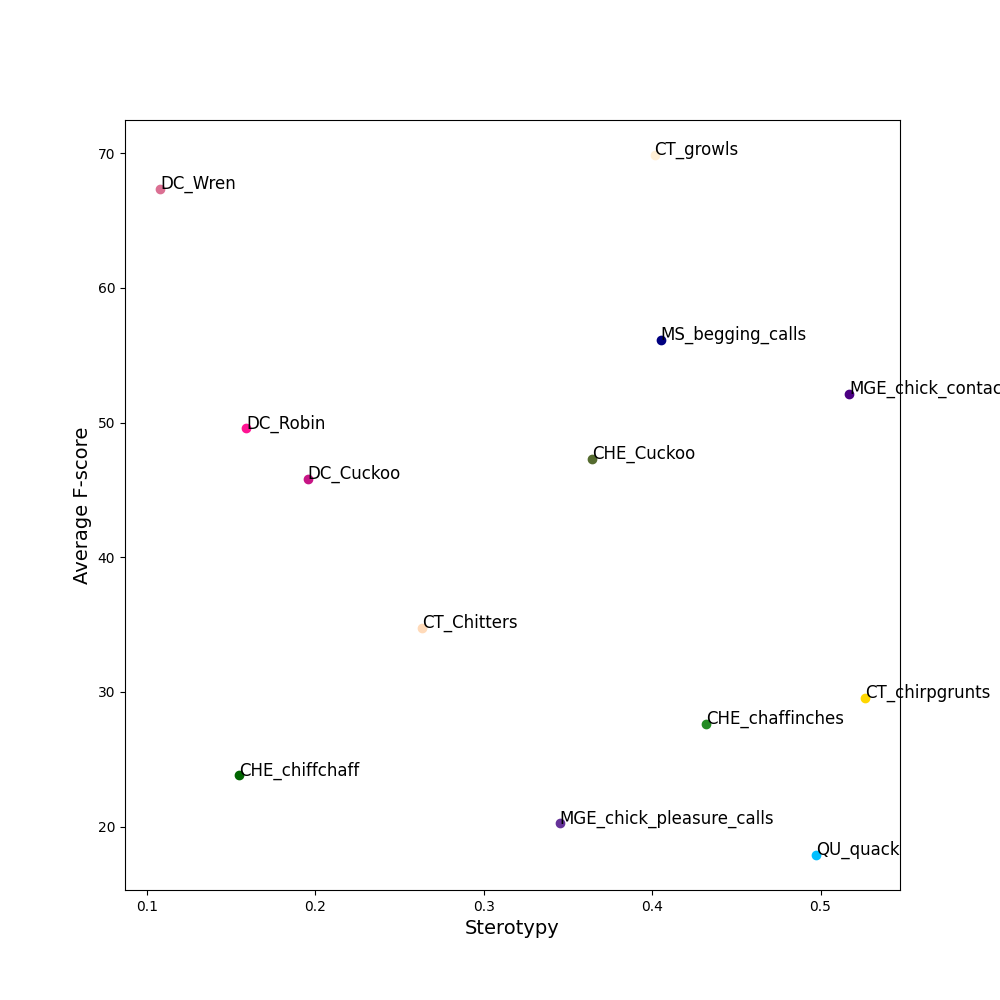}}
{\includegraphics[width=0.4\textwidth]{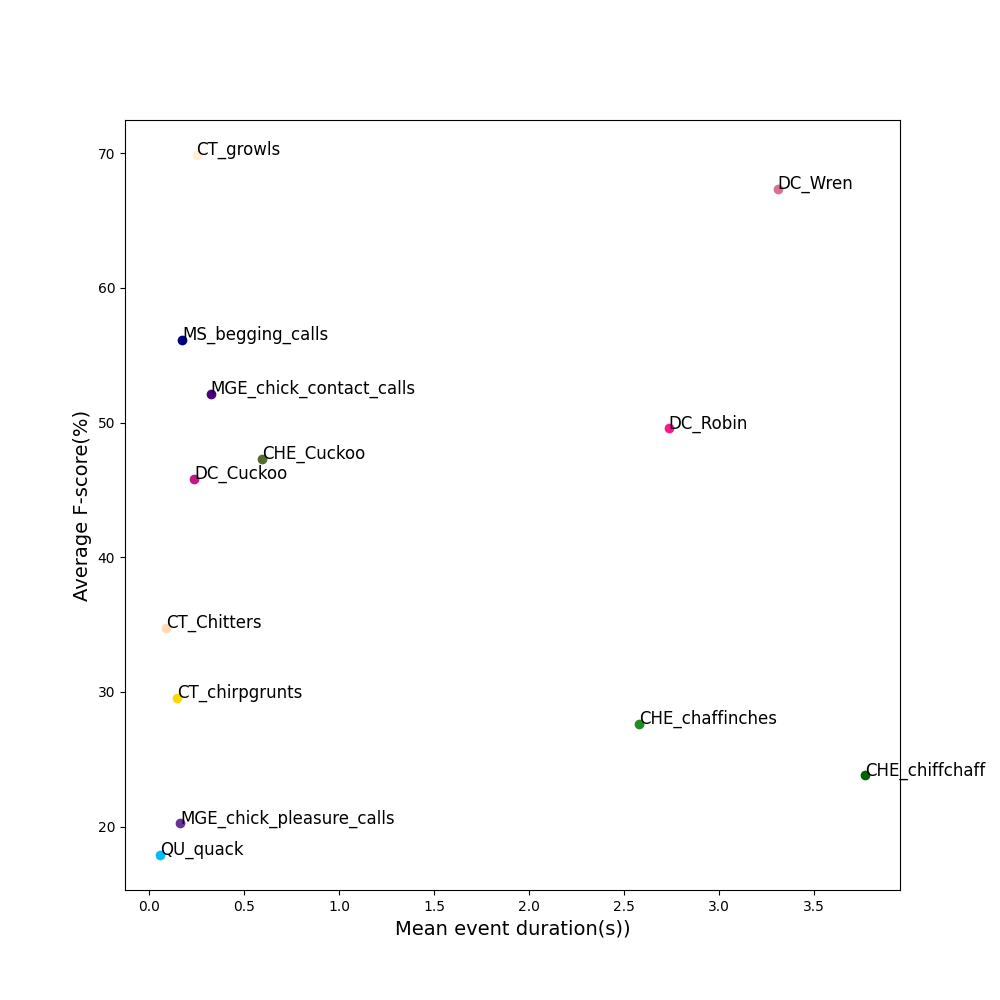}}
{\includegraphics[width=0.4\textwidth]{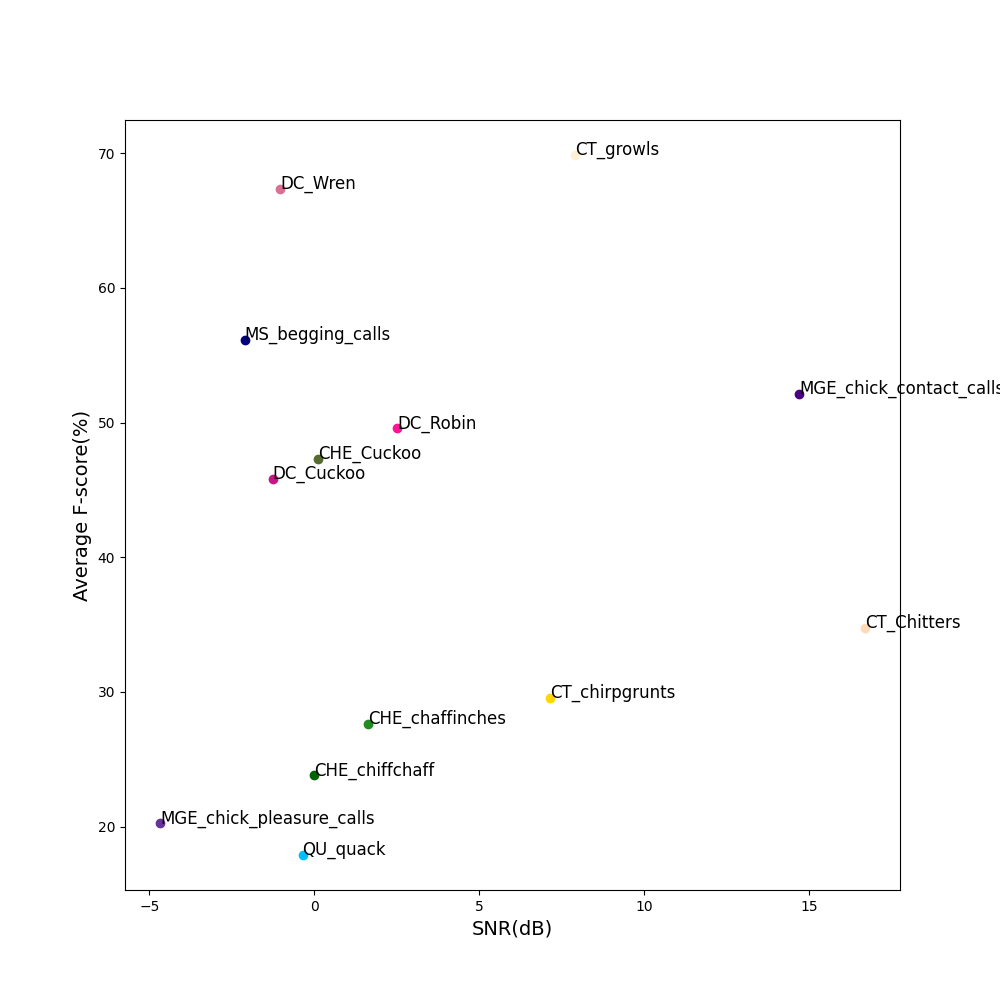}}
{\includegraphics[width=0.4\textwidth]{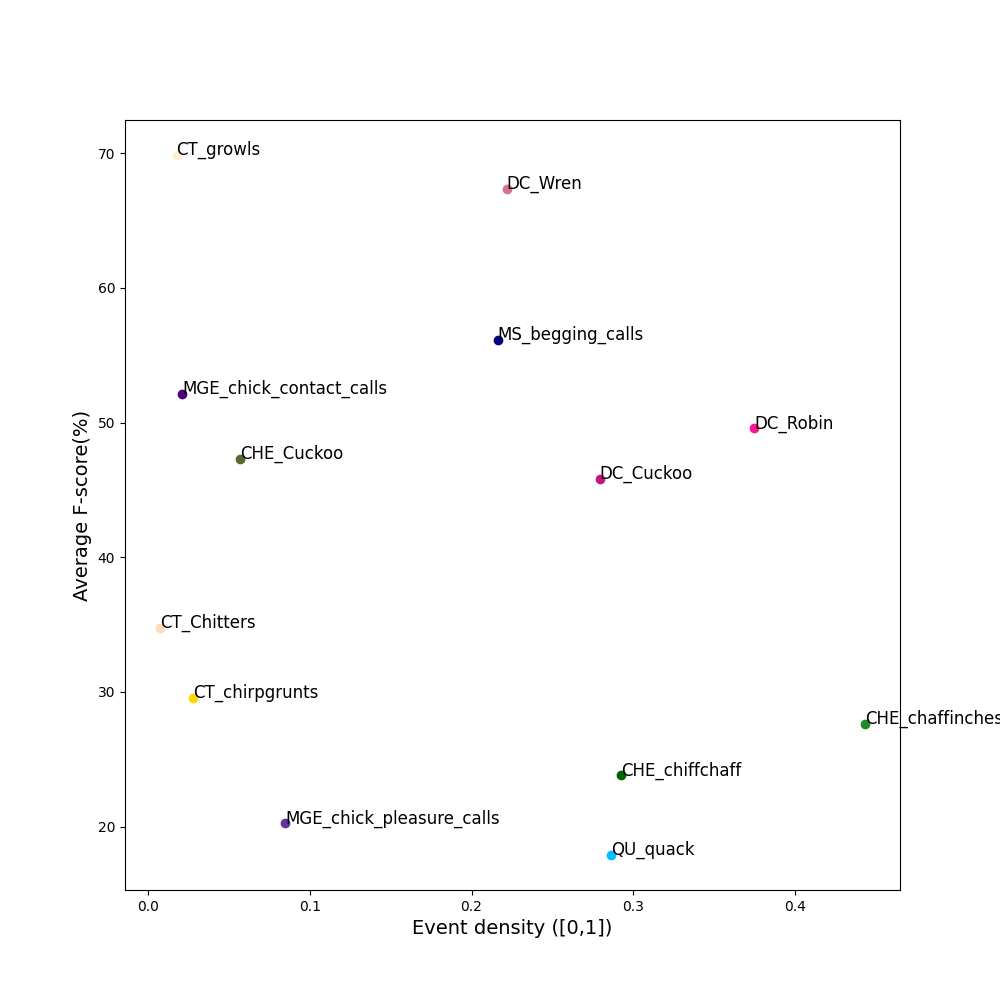}}
\caption{Scatter plots illustrating the relationship between several data characteristics and the F-score for each class of the evaluation set. Note the vertical axis represents an averaged F-score, which is calculated from all the systems scoring above the baseline. The horizontal axis displays five factors, namely similarity to 5 shots, stereotypy, event length, event density and signal-to-noise ratio (SNR).}
\label{fig:scatterplotFscoreDatacharacteristics}
\end{figure}

To determine which data characteristics might be the strongest factors in these performance variations, we investigated five data attributes, three commonly considered in soundscape analysis: SNR, event sparsity, and event length, plus similarity between events and the 5 shots and stereotypy (as defined in section \ref{sec:datasets}).
We performed a multivariate regression with different combinations of these variables. By evaluating and selecting the best model, we can verify which would be the best attribute or combination of attributes that predicts the Fscore.
The possible 31 combinations of these attributes were used as the predictors of the average F-score across all systems scoring above the baseline. The resulting regression models were then evaluated by inspecting the p-values, adjusted R-squared, Akaike information criterion (AIC) and the Bayesian information criterion (BIC)
The results indicate that none of these factors translating bioacoustic considerations was a strong predictor of differences in performance. The same can be observed in Figure \ref{fig:scatterplotFscoreDatacharacteristics}. 



%

\subsection{Ablation study }
The developed systems are complex and most consist of various independent functional units coming together to solve the task. 

Here, we present the results of the ablation study performed on Liu\_Surrey's system \citep{Liu:2022a}. 
An ablation study consists in removing different parts of the network and evaluating the impact these changes have on performance. This allows for some increased understanding of how a system works, while providing a way in which it is possible to measure the contributions of each individual unit for the overall level of performance achieved.

The experiments with variations to the system's architecture can be organized into different categories: 1) exploring different input features, 2) analysing the impact of Contrastive Learning and 3) impact of the number of ``ways'' used for episodic training ((as in the Meta-learning setup N\_way, K\_shots): ``ways'' means the number of different sound categories considered at once). The F-score results are presented in Table \ref{tab:ablation_study}.
Because systems are developed based on the development and validation sets alone, the design decisions might not be what works best for the evaluation set, specifically in the case where these datasets vary substantially. 
This explains why the original submitted architecture (first row of table \ref{tab:ablation_study}) did not result in the best performance across all the variations tested here. 
In fact, instead of applying \gls{PCEN} and Delta MFCCs as input features, the system modified to use Log Mel spectrograms resulted in the top performance on F-score.
Similarly, here we see that including Negative Contrastive learning does not work well on this evaluation set and indeed the performance of the system decreases.
Experimenting with different number of ways, confirms the expected that as the number of ways in the support set increases from 10 to 30, so does the performance. 
Finally, ensembling all the variations of (B) leads to an improved F-score. 

\begin{table}[]
\centering
\begin{tabular}{l|c}
\textbf{System Variation           }                                      & \textbf{F-score }     \\ \hline
PCEN+DeltaMFCC (original submitted system (B))                                 & 35.019          \\ \hline
only PCEN                                                                           & 36.001           \\
\textbf{only LogMel}                                                                & \textbf{40.355}  \\
LogMel+DeltaMFCC                                                               & 37.518           \\ \hline
w/o. Negative contrastive learning                                             & 39.637           \\  \hline
\#ways 10                                                                             & 35.503        \\
\#ways 20                                                                             & 35.866          \\
\#ways 30                                                                             & 36.608            \\
\#ways 40                                                                             & 36.021       \\  \hline
\textbf{Ensemble all} &
\textbf{50.624} \\
\end{tabular}
\caption{F-Score results for the different system variations on the evaluation set. First row refers to the unchanged submitted system, all the other systems are simple modifications to this.}
\label{tab:ablation_study}
\end{table}

\subsection{Expert use analysis}

We are interested in understanding how far away the best scoring systems are from being incorporated into the annotation practice and how helpful or misleading their predictions can be.
The expert annotators of QU, CT and MS datasets were given the predictions resulting from the 3 top scoring systems (\cite{Tang:2022, Liu:2022a, Martinsson:2022}) and asked to analyse them in terms of a) Usability and  b) Types of errors. Here are the main topics and highlights received. The full feedback can be read in \ref{app:expertanalysis}

All consider that at least one of the systems results in useful predictions that can be used as a starting point for manual editing.

The best ranking system overall (\cite{Tang:2022}) is not always the one selected by the experts as the best predictor of events in the different datasets and it also changes for different classes within the same dataset. However the experts' selection almost always agree with the F-score results by class shown in Figures \ref{fig:Fscore_dataset2022} and \ref{fig:fscores_by_class}.

As to the type of errors, the experts identified several instances of missed detections, misclassifications either on non-target calls or noise events, and in general imprecise detection of the duration of calls. 

Another aspect highlighted for both QU and CT datasets were situations where the capability of the systems to produce correct predictions decreased over time, meaning that events happening further away from the beginning (where the 5 shots examples happen) were less well predicted. 

The reason for some missed detections might be due to the selection of the 5 examples from which the systems need to learn the pattern of the target class. For both MS and CT datasets the experts commented that the range of variation of the target calls was not well captured within the 5 initial examples. This aspect is also expressed in Figure \ref{fig:similaritywith5shots}.  

Finally the potential for using FSL to improve upon human manual annotations is illustrated in the feedback received for the CT dataset. The system ranked in second place overall, \cite{Liu:2022a}, was able to predict 20 new Growls that the human annotator had not identified. 

\section{Discussion}


In this work we have formulated few-shot bioacoustic event detection as a machine learning task. We have evaluated many approaches to the task, and demonstrated that both meta-learning and transfer learning methods can successfully generalise \gls{FSED} to novel sub-tasks in bioacoustics---thus, transcribing animal sounds with a precision unobtainable with traditional methods, in the absence of huge training datasets.
Our sub-tasks were chosen to be diverse and non-trivial: they differed in taxon, target sound characteristics, background noise, stereotypy, stationarity, duration, and more.
We believe that we have shown that the many related recognition tasks in computational bioacoustics can be unified within a generalised approach to machine learning.

Leading systems achieve over 30\% F-score on all 6 tasks. This is a dramatic improvement over classic template-matching, and also over a standard modern deep learning approach (both of which often achieved F-scores below 10\%, in our baseline implementations). This reflects the fact that most bioacoustic sound event detection tasks have unique characteristics (such as noise, non-stereotypy, distractor sounds, non-stationarity) which make them distinct from each other and very hard to analyse with a conventional detection system.
Although automatic detection has been in use for many years, it has often required manual tweaking of a system's parameters for each new situation. 

Based on this study we believe our formulation of FSED is a useful one.
It is applicable across a wide selection of bioacoustics tasks, and provides a good target for machine learning development. It is not trivially solved by prior art in few-shot learning, nor by pretrained networks; yet we report very strong progress through the public challenges.
We also consider that our chosen evaluation measure---an event-based F-score---has good external validity, since it aligns well with expert evaluations of automatic transcripts.

Our aim to generalise over a range of loosely-related datasets/tasks is of current interest in machine learning.
There are some comparable initiatives in wildlife monitoring.
The `BEANS' project collects together animal sound datasets and aims to provide a general evaluation benchmark \citep{Hagiwara:2022}. Their work focuses on classification rather than temporal detection, and it does not consider few-shot learning or meta-learning---however it may be possible to re-use their data for such things.
Similarly, in image recognition the NeWT benchmark provides a suite of tasks for wildlife images, using a classification framework to ask a very wide range of ecology questions from a single data representation \citep{vanHorn:2021}.
Key differences between these and our work are our few-shot setting, and the explicit inclusion of temporal structure in the input and output of our formulation.

Based on the results presented here, are few-shot bioacoustic SED systems good enough to use? Yes, as determined by feedback from our panel of experts. Although the outputs from such systems are far from perfect, they were judged to be of sufficient quality for active use, in place of fully-manual annotation.
The quantitative results demonstrate that, when presented with a new dataset with no large training corpus, few-shot bioacoustic SED outperforms common methods such as template-matching, as well as standard transfer learning from well-known large datasets (AudioSet, BirdNet).
It is worth remembering, however, that our paradigm is designed for the case of detecting events for which no large training dataset is available. If large amounts of labelled data are available, or pre-trained networks whose training matches well with the intended use, then the more common machine learning method (i.e.\ supervised learning) would be expected to be the most reliable approach.

\subsection{Aspects of bioacoustic datasets that affect performance}

Many aspects of bioacoustic datasets make them complex to analyse: noise, highly sparse or dense events, varied levels of stereotypy, and non-stationarity (drift) in conditions.
We selected datasets which varied across many of these characteristics, and we sought to evaluate which of them were key factors influencing the difficulty of the task.
A quantitative analysis (multivariate regression) was unable to identify any factors that consistently affected the F-score results across these datasets.
However, qualitative feedback from expert users indicated that non-stationarity exerted an effect: this was shown by a reduction in performance for time-regions distant from the annotated examples.

A separate issue picked up by our annotators was that the support set was not always a good representation of the class 
to be detected. either because it did not include examples of every call type or due to the low stereotypy characteristics of certain classes.
A trivial response is that we may include more than 5 examples, or curate the examples so they span the desired range of calls.
In such cases we would expect stronger performance, at the slight cost of reintroducing some manual intervention.
Our design decision to use the \textit{first} few examples in a dataset is reflective of initialising a system before deploying it in a new recording situation. With offline analysis, and more flexibility in selecting examples, higher performance can be achieved.

Since bioacoustic targets may include multiple call types or non-stereotyped sounds, it is worth noting one aspect of prototypical networks.
The formation of a single fixed prototype, by taking an average in a coordinate space, implies that the examples are in some sense all of one kind.
This assumption is also challenged by non-stationarity, which we might think of as a prototype gradually drifting rather than remaining fixed.
Transductive inference helps to reduce these issues by allowing the feature space to be updated at query time, and this was used by various strongly-performing systems.
There remains much opportunity for multiplicity and drift to be included into designs for the \textit{concept formation} of \gls{FSED} systems.

\subsubsection{Duration of animal sound events}

The annotated durations of animal sound events can range over multiple orders of magnitude, from milliseconds to minutes.
They result from diverse physical processes, from the impulsive (dolphin clicks) to the continuous (mosquito flight).
In retrospect it is clear that this needs to be handled carefully in the design of an SED system, because many computational methods have inbuilt assumptions or limitations in the durations they can process.
When at first we formulated the task, we did not foresee that correct handling of event durations would be an important factor in evaluation performance, but this was indeed the case.
The variable scale of event durations is not a limitation in itself---the difficulty comes when we try to solve all these different tasks with very different characteristics, together in one algorithm.

Many machine learning systems have pragmatic design constraints that limit the range of durations they can consider. Our template-matching method uses ranges directly inherited from the 5 annotated events, although there remain practical limits on very large templates, such as computer memory.
In deep learning, long audio files are usually divided into shorter chunks (with fixed durations of e.g.\ 3 or 10 seconds), so that they can fit inside the limited memory of GPUs. To detect long events, detections that span these chunks are joined together in post-processing.
This as well as other considerations meant that post-processing of outputs was an important aspect of all strongly-performing systems.

The (deep) feature extraction procedure itself can place limits on event durations. Firstly the resolution of spectrograms, with a typical granularity around 10~ms per `frame', often predetermines the finest scale that can be resolved.
Datasets QU and MGE contained very short sounds at around this scale; \citet{Tang:2022}'s framewise CNN excelled on these datasets.
Secondly, CNNs (used by all submitted systems) have relatively small ``receptive fields'' and do not consider the whole spectrogram but local feature patterns.
Many of the strongest systems adapted themselves at query-time to the expected event length inferred from the 5 examples, in particular \citet{Martinsson:2022} who trained a set of embedding functions, each designed for a different duration.
Some notable systems augmented their core CNN with architectures that are able to integrate information over long durations and directly infer onset and offset locations, such as a CRNN event filter \citep{Tang:2022}, perceiver and/or region proposal network \citep{Wolters:2021}.
We envisage that future developments on these lines may be fruitful, perhaps using further techniques from object detection.


%
%
%
%
%
%
%

\subsection{A single method for bioacoustic SED?}

Our baseline prototypical network is itself novel since \gls{FSL} has been applied to sound event classification, but almost never (prior to our work) to SED.
Through a public data challenge we have seen many different variations on this method, leading to strong results.
Is it possible to recommend a single method to take forward for bioacoustic SED; and if so, does it use prototype-based meta-learning?

We find that prototype-based meta-learning works well when taking care about certain aspects of the method (namely the choice of negative examples, and duration filtering / postprocessing of events).
Conversely, the strong performance of some non-prototypical systems \citep{Yang:2021,Tang:2022,Wu:2022} indicates that the paradigm is not a necessary component.
Bioacoustic FSED can be addressed by either meta-learning or fine-tuning approaches.

Query-time adaptation (transductive inference) was shown in multiple cases to lead to very strong performance, within both the prototypical and fine-tuning paradigms. This comes at a cost of added complexity and added query-time computation, since typically a new run of statistical optimisation must be performed for a new query task.
Thus, from the present results we can recommend that a system should include query-time adaptation for the best possible detections, but that a system without query-time adaptation should be a widespread default. Such fixed embeddings can easily be used off-the-shelf, in the same way that other pretrained networks are now commonly downloaded and used.
The DFSL method employed by \citet{Wu:2022} is an alternative approach which combines an unchanging feature extraction with a query-time adaptive weighting. This combines stability with dynamic adaptation, and thus is worthy of further investigation.

\subsection{A single embedding for bioacoustic SED?}

Contrary to query-time adaptation, in machine learning there is current interest in learning good feature representations (good embeddings) from data.
If an embedding can be re-used unmodified, this has an appeal of providing a general, reusable, and potentially low-complexity analysis tool, a component to be used in many systems.
For audio data, some of the most widely-used deep embeddings are those derived from pretraining with the large-scale \textit{AudioSet} dataset, originally designed for classifying many different (human-centric) acoustic categories \citep{Gemmeke:2017}.
More recent work evaluates this and many more ways to create an embedding \citep{Turian:2022}.

Our evaluation shows that improved prototypical network methods create powerful embeddings, useful even with no test-time adaptation. It is impressive that a single vector space could be used to represent our diverse bioacoustic tasks.
The present work on few-shot learning thus offers a different perspective on representation learning for sound in general, and animal sound in particular.

We believe there is yet more potential to these approaches in creating a generalisable single solution for the many varied bioacoustic tasks.
With that purpose, the challenge is running a third time in 2023. Although the setup remains the same from previous editions, the evaluation set has been extended and a new rule is introduced - Ensemble models are no longer allowed. 
This way we are pushing the systems to cover more bioacoustic tasks, and to do so as a single generalisable model.

Finally, it is possible to start envisioning the implementation of such systems for practical use. Some of the challenges identified here, due to the selection of the 5 shots or the non stationarity of long audio recordings, can easily be addressed.
This paradigm thus moves us towards a post-template matching era for bioacoustic sound event detection.

\section*{Author contributions}


Conceptualization: DS, VL, ASP

Methodology: DS, VL, ASP, LG

Software: VM, IN, SS, VL

Validation: VM, IN, SS, LG

Investigation: VM, IN, SS

Data curation: ASP, LG, HP, HW, IK, FJ, JM, ME, EV, EVV, EG, IN

Writing - Original Draft: DS, VL, VM, IN, SS, ASP, LG, HP, HW, IK, FJ, JM, EV, EVV

Writing - Review \& Editing: DS, VL, ASP, HW, LG, IN

Visualization: EVV, IN, SS

Supervision: DS

Project administration: DS

\section*{Acknowledgments}

ASP acknowledges funding from Human Frontier Science Program award RGP0051/2019. The work was also funded by the Deutsche Forschungsgemeinschaft (DFG, German Research Foundation) under Germany's Excellence Strategy—EXC 2117—422037984. ASP received additional support from the Gips-Schüle Stiftung and the Max Planck Institute of Animal Behavior. 

The data for meerkats were collected at the Kalahari Meerkat Project in South Africa, currently supported by a European Research Council Advanced Grant (No. 742808 and No. 294494) to Tim H. Clutton-Brock, the MAVA Foundation, and the University of Zurich. We further thank the Trustees of the Kalahari Research Centre and the Directors of the Kalahari Meerkat Project.

Spotted hyena data were collected in collaboration with the MSU-Mara Hyena Project, and data collection was additionally supported by a grant from the Carlsberg Foundation to FHJ. 

\clearpage
\bibliographystyle{plainnat}
\bibliography{refs}

\clearpage

\appendix

\section{Supplementary info}

\subsection{Measuring similarity between events}
\label{app:sterotypy}

A metric to evaluate similarity between sound events is needed in order to analyse two aspects of the datasets: 1) How well do the initial 5 POS events represent the remaining POS events and 2) How stereotyped are the vocalisations in each dataset.\footnote{\url{https://github.com/inesnolas/acoustic_stereotypy}}

Here, similarity between two events is defined by the maximum value of their cross correlation. i.e : 
\begin{equation*}
    sim(t,e) = max_{k} [xcorr(stft_t, stft_e(k:k+L))] 
\end{equation*}

 where $stft_t$ is the short term fourier transform of the template event(STFT), and $stft_e(k:k+L)$ is a slice of the STFT of a POS event e; k being the starting time index and L being the duration of the template event t in STFT frames.

Both procedures to compute 1) and 2) are similar and based on averaging the similarity of randomly selected events. 
The first step consists in selecting the "template" events: in 1) these are the initial 5 POS events while in 2) these are a random selection of 10 POS events across the whole audio recording.
Each of the template events is then cross-correlated with 30 randomly selected POS events. The average of the maximum cross correlation across the 30 operations results in a single value representing the average similarity between each template event and the remaining POS events in the audio file.
The final step is to average again this similarity value across all templates. Formally, these operation can be written as: 
\begin{equation*}
     \frac{1}{T}\sum_{t}^{T}\frac{1}{E}\sum_{e}^{E}sim(t,e), 
\end{equation*}
 where T is the number of template events (either 5 or 10) and E is the number of POS events randomly selected (30 in this implementation)
This proposed metric to measure similarity presents some limitations, namely events that differ from the templates on the time domain will be overly penalized, while a human annotator might still consider them to belong to the same class. A common example is when events present a similar pattern except that they differ in duration or because they are time-stretched.

Finally, when comparing stereotypy values across different classes, it is important to note the different granularity that these labels represent. As it is expected classes representing a specific call type or even calls from a single individual should have higher stereotypy values than broader classes. The results of these comparisons across different datasets are thus limited to the purpose of assessing the characteristics of the different datasets.




\pagebreak

\subsection{Expert analysis of predictions}
\label{app:expertanalysis}

Expert analysis of the predictions produced by the overall top 3 ranking systems.
For this analysis we asked the experts who annotated the data for CT, MS and QU datasets to answer the following questions and provide general feedback on how well the systems did in their specific datasets.

\begin{enumerate}
    \item Usability of the predictions as a tool. Are the predictions good enough to use without any manual editing? If not: what would be the relative time cost of editing the predictions, versus starting again with manual annotation? Could these outputs, as they are, facilitate your work?
    \item Error analysis. In what ways does it go wrong? (e.g. too many false positives; onset/offset times inaccurate; sound events become split apart or merged together.) By inspecting the data, what seems to cause errors? (e.g. moments of high background noise; calls from other animals; non-stereotyped calls missed; conditions changing.) And are there any obvious ways that you think would correct these errors? (selection of 5 different shots?, segmenting the audio files in shorter sections?)
\end{enumerate}

\subsubsection*{\textbf{MS dataset (Manx Shearwaters)}}

Feedback by JM:
(answers to the questions above for each of the systems independently)

\begin{itemize}
    \item Du\_NERCSLIP
        \begin{enumerate}
            \item Good. The predictions successfully classify the target class of chick begging vocalisations. Additionally, they rarely misclassify adult grunting vocalisations or other background sounds as chick begging vocalisations. Editing the predictions would be quicker than starting again with manual annotation. These predictions could facilitate our work.
            \item The errors, in most cases, are missing out instances of chick begging vocalisations. In particular, fast bouts of begging are in some cases missed entirely, or only a small subset of chick begs are classified. The five shots from the beginning of the file do not come from fast begging bouts, and so are not representative of the range of possible chick begging vocalisations. Additionally, the onset and end of predictions are typically imprecise, with the onset often slightly early; in fast bouts the onset and end of begs is particularly imprecise. The error profile is relatively similar to Liu\_Surrey.
        \end{enumerate}
    \item Liu\_Surrey
        \begin{enumerate}
            \item Good. The predictions successfully classify the target class of chick begging vocalisations. Additionally, they only occasionally misclassify adult grunting vocalisations or other background sounds as chick begging vocalisations. Editing the predictions would be quicker than starting again with manual annotation. These predictions could facilitate our work.
            \item The errors, in most cases, are missing out instances of chick begging vocalisations. In particular, fast bouts of begging are in some cases missed entirely. The five shots from the beginning of the file do not come from fast begging bouts, and so are not representative of the range of possible chick begging vocalisations. Additionally, the onset and end of predictions are typically imprecise, with the onset often slightly early; in fast bouts the onset and end of begs is particularly imprecise. Adult vocalisations and background noise are occasionally misclassified as chick begging vocalisations. The error profile is relatively similar to Du\_NERCSLIP.
        \end{enumerate}
    \item Martisson\_RISE
        \begin{enumerate}
            \item Excellent. The predictions successfully classify the target class of chick begging vocalisations. Sometimes they misclassify adult grunting vocalisations or other background sounds as chick begging vocalisations. The onset and end of chick begging vocalisations are identified precisely in many cases. Editing the predictions rather than starting again would be much quicker. These predictions certainly could facilitate our work. 
            \item The errors, in many cases, are missing out instances of chick begging vocalisations. In particular, fast bouts of begging are in some cases missed entirely. The five shots from the beginning of the file do not come from fast begging bouts, and so are not representative of the range of possible chick begging vocalisations. Additionally, adult vocalisations and background noise are sometimes misclassified as chick begging vocalisations. The error profile of Martinsson\_RISE differs from the error profiles of Du\_NERCSLIP and Liu\_Surrey. 
        \end{enumerate}
\end{itemize}

\subsubsection*{\textbf{CT dataset (Coati)}}

Feedback by EG:

General answers to question 1) and 2) above: 
\begin{enumerate}
    \item The usability of the predictions is dependent on the call type under detection. The chitter predictions from H1 were best performing because they found at least one chitter in most of the chitter bouts, so manual labelling for the other chitters in these bouts would be necessary. A 1-hour wave file with many chitters can take 8 hours of manual labelling – so having a tool to pinpoint the areas to focus labelling effort would save time, it would likely save 2-4 hours of manual labelling time which would facilitate our work. D1 and M1 missed most chitters so I would not use these for labelling. In general, I prefer over-predicting calls to under-predicting, as deleting incorrect labels is faster than listening to whole wave files. 
    
    The growls were best predicted by H2, as they found 20 more growls which were faint to the human labeller and therefore missed. These labels would still need editing as the call durations were longer than the actual call, but I was impressed at its detection capabilities. D2 was the second best at detecting calls and M2 was the worst (missing 22 growls). 
    
    For chirpgrunt detection, D3, H3, and M3 were similar in performance but the call durations varied between them. The chirpgrunt durations were best predicted by D3, however 6 calls were missed (likely because of increased background noise and the chirp component was fainter). H3 missed the least chirpgrunts but the duration of the calls was longer – which would take manual correction. H3 also mislabelled a bird call for a chirpgrunt. M3 durations were shorter for 22 chirpgrunts, so the grunt component was missed, this would also need manual correction.

    \item For the chitter predictions, different shots should be used which better represent the variation of chitters (in frequency/amplitude/duration). For the chirpgrunts and growls, the missed labels were when the chirps/growls were fainter or there was background noise. Again, I would give more varied chirpgrunts/growls in the shots to account for this variation. I noticed that the growl predictions were less accurate over time, so shorter segments may also increase the accuracy. 
    
\end{enumerate}

Comments about the selected 5 shots:

 \begin{table*}[h]
     \centering
     \begin{tabular}{c|p{0.8\textwidth}}
          Class &  \\ \hline
          Chitters &  Training events were faint and not good examples for the classifiers, which I think heavily affected the quality of the events for this call type. The call shape, frequency and amplitude of chitters are highly variable – so having a range of different chitters may make the classifiers more accurate. These calls are also usually emitted rapidly in bursts of around 4 to 20 calls depending on the severity of the aggressive interaction\\ \hline
          Growls & Training events were good examples for classifiers, but call durations were not that varied which may affect the classifiers duration of calls \\  \hline
          Chirpgrunts & Training events were good examples for the classifier \\ \hline
     \end{tabular}
     \label{tab:my_label}
 \end{table*}

Observations on each system's predictions for each audiofile/class: 

\textbf{Du\_NERCSLIP predictions on ct1.wav (chitters):}
\begin{itemize}
    \item 6 chitters were found (out of 99) which were lower in frequency to the “average” chitter, but these were more similar to the training labels
    \item no mislabeled chitters
    \item better duration accuracy than M1
\end{itemize}

\textbf{Du\_NERCSLIP predictions on ct2.wav (growls):}
\begin{itemize}
    \item 1 growl was split into 2
    \item 4 growls found which was not in gt 
    \item louder growls were labelled shorter than call length
    \item 3 mislabeled growls for background noise
    \item 2 growls missed which were during chittering bouts (not in training calls)
    \item 4 growls missed (unclear why)
    \item 3 growls missed which were shorter in duration to training calls
    \item calls were less accurately labelled by end of file
     \end{itemize}

\textbf{Du\_NERCSLIP predictions on ct3.wav (chirpgrunts):}
    \begin{itemize}
    \item duration of labels is similar to gt labels
    \item 6 chirpgr labels missed – the chirp component in these calls were quieter and there was more background noise. for one of these mislabels, the chirp was in a higher frequency to the training data
     \end{itemize}

\textbf{Liu\_SURREY predictions on ct1.wav (chitters):}
    \begin{itemize}
    \item more chitters were labelled compared to D1 but the durations were roughly double the length of the call
    \item mislabeled bird calls for chitters (they are similar in call duration and shape)
    \item shorter chittering bouts were more accurately labelled than the longer bouts
    \item at least one label in each chittering bout which is helpful to locate these bouts, but these calls would need to be manually relabeled 
     \end{itemize}

\textbf{Liu\_SURREY predictions on ct2.wav (growls):}
    \begin{itemize}
    \item some of the call durations were longer than the call
    \item 20 growls labelled which were not in gt (they were much fainter)
    \item 13 growls mislabeled – actually background noise (5 were chirpgrunts)
    \item 5 mislabeled – actually grunts (shorter in duration)
    \item overall I was impressed with these labels, would need some corrections but it was able to pick up faint growls better than a human (perhaps because they are harder to hear at the lower frequencies?)
     \end{itemize}

\textbf{Liu\_SURREY predictions on ct3.wav (chirpgrunts):}
    \begin{itemize}
    \item all labels start and end longer than the call duration so this would need to be manually corrected (which would take some time)
    \item missed chirpgr where the chirp was in a higher frequency to the training data
    \item also missed 3 labels that didn’t have obvious differences in the call amplitude/quality/frequency to the training data
    \item mislabeled bird call for chirpgr
     \end{itemize}

\textbf{Martinsson\_RISE predictions on ct1.wav (chitters):}
    \begin{itemize}
    \item missed most of the chitters (7/99 found), the chitters labelled were more similar to training data
    \item similar results to D1, no mislabeled chitters 
    \item length of chitters longer than gt labels 
     \end{itemize}

\textbf{Martinsson\_RISE predictions on ct2.wav (growls):}
    \begin{itemize}
    \item 22 growls missed (these growls were much fainter and some had background aggressive calls)
    \item 2 grunts mislabeled at growls
    \item overall under labeled compared to D2 and H2 
     \end{itemize}

\textbf{Martinsson\_RISE  predictions on ct3.wav (chirpgrunts):}
    \begin{itemize}
    \item 22 of the labels end before grunt component of call, so would need correcting
    \item missed chirpgr where the chirp was in a higher frequency to the training data
    \item mislabeled background noise for chirpgr
    \item missed 5 chirpgr where the chirp component was fainter/more background noise
    \end{itemize}

\subsubsection*{\textbf{QU dataset (Dolphin Quacks)}}

Feedback by FJ:

Overall, Du\_NERCSLIP by far is the best and actually seems relatively useful. Something is weird with the first file where somehow it did not catch much. For the other files, performance is generally quite good, with fair bit of misses and occasional merged, and sometimes bounds are a bit wide too. However, this one could certainly be used as a starting point where manual revisions could then fix potential errors, and I think it would save a lot of time in this way. I was particularly impressed by how robust this one was to different noise conditions, including loud vessels and also LOTS of other dolphin distractor sounds, many of them very loud and overlapping the target sounds. Sometimes it seemed to be triggered on pulsed signals that were not the target category but that did  not seem to be always and may depend on characteristics of the 5 known signals.

Liu\_SURREY performance was relatively poor, subjectively speaking. For the first three files it triggered near-consistently before the actual signal, with both start and end bounds in the gap between signals rather than covering signals. That was not true for some of the subsequent files - wonder if there is a risk that bounds were exported with a negative delay somehow. In general this one tended to have lots of false detections, especially *at least for a few files where I noted it) broadband short pulsed distractors. For a few files it also seemed like performance deteriorated over time but this did not seem consistent.

Martinsson\_RISE seemed extremely conservative, with several files without any detections at all, and mostly misses or triggering on noise rather than correct detections.

\clearpage
\subsection{Few-shot task 2021}

The main results shown in the paper relate to the 2022 edition of the few-shot challenge, organised as a task within DCASE 2022.
That was the second edition. Here, we show the results of the first edition of the challenge (2021), which was very similar in design but with fewer datasets.

The datasets are described in Table \ref{tab:datasets2021},
and characteristics of the submitted systems as well as their reported performance in \ref{tab:teams2021characteristics}.
Most of the 2021 datasets were reused in 2022, although with some datasets expanded or annotations corrected.

In addition to datasets described in the main text, the 2021 edition included one dataset labelled ML, using 17 recordings extracted from the Macaulay library. Each recording contains calls from a different species: 14 terrestrial mammals (not including hyenas or meerkats) and 3 birds (not including passeriformes).
The Macaulay Library is a digital archive of images, videos, and sounds from animals.\footnote{Official website: \url{https://www.macaulaylibrary.org/}}
As of 2021, it contains 175k audio recordings from 10k species of birds and 2k species of amphibians, fish, mammals and insects.
These recordings are contributed by amateur and professional recordists around the world, and the catalogue is maintained by the Cornell Lab of Ornithology.
For the DCASE 2021 challenge, one author (DB) curated 17 recordings from the Macaulay Library and annotated them in terms of animal vocalisations.
The average duration of each recording is of the order of one minute and the number of calls per minute varies in the range 10--150.

The ML dataset was used in the 2021 evaluations; however, for the 2022 it was withdrawn after finding that the annotations were not of sufficient temporal precision.

\begin{table}[ht]
    \begin{tabular}{r|c|c|c|c|c|c|c}
          &  &  &  &  \# & total & \# & \#  \\
         \textbf{2021} & Dataset & Taxon & mic type &  files & duration & labels & events \\
         \hline
        \multirow{4}{*}{Training}
        & BV & Birds & fixed & 5 & 10 hours & 11 & 2,662\\
        & HT & Mammals & on-body & 3 & 3 hours & 3 & 435\\
        & MT & Mammals & on-body & 2 & 70 mins & 4 & 1,234\\
        & JD & Birds & on-body & 1 & 10 mins & 1 & 355\\
        \hline
        \multirow{2}{*}{Validation}
        & HV & Mammals & mobile & 2 & 2 hours & 2 & 50 \\
        & PB & Birds & fixed &  6 & 3 hours & 2 & 260 \\
        \hline
        \multirow{3}{*}{Evaluation}
        & ME & Mammals & handheld & 2 & 20 mins & 2 & 70\\
        & ML & Mammals/birds & various & 17 & 20 mins & 17 & 1,035 \\
        & DC & Birds & fixed &  13 & 105 mins & 3 & 967\\
         
    \end{tabular}

\caption{Information on each dataset. Note that most datasets from 2021 were reused in 2022, though some datasets were expanded (HT) or received corrections to annotations. Subtotals are calculated excluding UNK.}
    \label{tab:datasets2021}
\end{table}

\subsubsection*{Results of the 2021 challenge}

The first public edition of this challenge in 2021 had 7 teams participating with a total of 24 submitted systems.\footnote{\url{https://dcase.community/challenge2021/task-few-shot-bioacoustic-event-detection-results}}
All submitted systems adopted prototypical networks (Table \ref{tab:teams2021characteristics}).
Data augmentation was applied by the majority of the teams, with SpecAugment being the most popular choice.
All systems relied on some sort of post-processing mechanism designed to remove superfluous predictions and many teams reported notable improvements in results due to such post-processing.
Another popular choice was using Per-channel Energy Normalisation (PCEN) \citep{Lostanlen:2019} as acoustic features.

The best ranked system improved over the baseline prototypical approach by applying a transductive inference method, where supplemental information is used to convey more representative prototypes of each category.
The system ranked in second place also improved over the prototypical baseline by using additional data from Audioset to train a ResNet for the feature extraction part. They have also adopted embedding propagation \citep{Rodriguez:2020}, with the objective of smoothing the decision boundaries as a way of increasing the generalisation capabilities of the few-shot system.

Also of note, the work in \citet{Cheng:2021} uses i-vectors as input features; both submissions in \citet{Zhang:2021} and \citet{Johannsmeier:2021}, explicitly create a negative class to model background noise and construct a negative prototype; and in \citet{Bielecki:2021}, the team opted for combining the prototypical loss, with knowledge distillation and attention transfer loss. 

For most high-performing systems, there was a drop in F-score from validation to the evaluation set (Table \ref{tab:teams2021}).
This suggests that the systems are generally dataset sensitive, and our datasets vary in difficulty.
To highlight this aspect further, we report the F-score results per dataset in the evaluation set  (Table \ref{tab:teams2021}), and also per-class (Figures \ref{fig:fscoreByClassDCset}
). Most systems have a low performance on the DC set, comprised of dawn chorus recordings, while perform better on ME and ML that include mainly mammal vocalisations. Complex acoustic environments such as dawn chorus may yet need further techniques to be employed for robust SED.

\begin{table}[t]
    \centering
    \begin{tabular}{r|l|c|c|c|c|c}
    Rank & Team name & Evaluation & Validation & DC & ME & ML\\ 
    \hline
    1    & Zou\_PKU  & \textbf{38.4}  (36.2 - 40.6)  & 55.3  & 20.6  &68.0 & 67.3 \\
    2    & Tang\_SHNU & \textbf{38.3} (36.1 - 40.5)   & 51.4 & 25.6  & 61.5  & 43.3 \\
    3    & Anderson\_TCD & 35.0 (33.1 - 37.0) & 26.2 & 19.9 & 56.6 &56.8  \\
    4    & Baseline\_TempMatch & 34.8 (32.6 - 37.1)  & 2.0  & \textbf{32.2} &47.1  & 29.5\\
    5    & Cheng\_BIT  & 23.8 (21.9 - 25.7)    & 46.3  & 10.6  & 53.5  & \textbf{78.8}  \\
    6    & Baseline\_PROTO  &  20.1 (18.2 - 21.9)  & 41.5 & 8.5 & \textbf{72.7}  & 55.7  \\
    7    & Zhang\_uestc  & 16.8  (15.5 - 18.2)  & 54.4  & 8.1 & 45.1 & 29.9 \\
    8    & Johannsmeier\_OVGU & 15.2  (13.7 - 16.7)& \textbf{58.6}  & 6.5  & 64.3  &35.8 \\
    9    & Bielecki\_SMSNG  & 8.4  (7.1 - 9.7)  & 51.8 & 3.1 & 56.3  & 51.4 \\
    \end{tabular}
    \caption{2021 F-score results (in \%) per team on evaluation and validation sets. Numbers in brackets indicate 97.5\% confidence intervals. The final three columns show the per-dataset scores for each evaluation dataset.}
    \label{tab:teams2021}
\end{table}

\begin{table}[tp]

\begin{tabular}{c|l|l}
\textbf{Rank} & \textbf{Team name} & \textbf{System characteristics} \\ 

\hline
1    & \begin{tabular}[c]{@{}l@{}} Zou\_PKU \\\citep{Yang:2021} \end{tabular}& \begin{tabular}[c]{@{}l@{}}CNN, \\Transductive inference, \\Mutual learning framework\\ Acoustic features: MelSpectrogram\\ Post-processing: peak picking, threshold.\end{tabular} \\ 
\hline
2    & \begin{tabular}[c]{@{}l@{}} Tang\_SHNU \\\citep{Tang:2021} \end{tabular}      
& \begin{tabular}[c]{@{}l@{}}ResNet,\\Prototypical Network\\ Embedding propagation, \\ Additional external data used.\\ Acoustic features: PCEN\\ Augmentation: SpecAugment, at inference time.\\ Post-processing: peak picking, median filtering\end{tabular}\\  
\hline
3    & \begin{tabular}[c]{@{}l@{}} Anderson\_TCD \\\citep{Anderson:2021} \end{tabular}   
& \begin{tabular}[c]{@{}l@{}}CNN,\\Prototypical Network  \\Acoustic features: PCEN, MelSpectrogram\\ Augmentation: SpecAugment\\ Post-processing: Probability averaging,\\ median filtering, minimum event length\end{tabular} \\
\hline
4    & Baseline\_TempMatch & Template Matching   \\
\hline
5    & \begin{tabular}[c]{@{}l@{}} Cheng\_BIT  \\\citep{Cheng:2021}  \end{tabular}    
& \begin{tabular}[c]{@{}l@{}}CNN,\\Prototypical Network \\Acoustic features: PCEN, i-vector\\ Augmentation: SpecAugment.\\ Post-processing: threshold.\end{tabular} \\
\hline
6    & Baseline\_PROTO  & \begin{tabular}[c]{@{}l@{}}CNN,\\Prototypical Network  \end{tabular} \\
\hline
7    & \begin{tabular}[c]{@{}l@{}} Zhang\_uestc \\\citep{Zhang:2021}  \end{tabular}
& \begin{tabular}[c]{@{}l@{}}ResNet,\\Prototypical Network\\Acoustic features: PCEN\\ Augmentation: SpecAugment\\ Post-processing: threshold.\end{tabular}  \\
\hline
8    & \begin{tabular}[c]{@{}l@{}} Johannsmeier\_OVGU \\ \citep{Johannsmeier:2021}  \end{tabular}
& \begin{tabular}[c]{@{}l@{}}CNN,\\Prototypical Network  \\ Acoustic features: PCEN, MelSpectrogram\\ Augmentation: Time stretching, Pitch and\\ Time shifting\\ Post-processing: threshold, gaussian smoothing\end{tabular} \\
\hline
9    & \begin{tabular}[c]{@{}l@{}} Bielecki\_SMSNG \\\citep{Bielecki:2021}  \end{tabular} & \begin{tabular}[c]{@{}l@{}}CNN,\\Prototypical Network \\ Knowledge Distillation and Attention transfer loss.\\ Additional external data used.\\ Acoustic features: MelSpectrogram\\ Augmentation: melspectrogram time and\\ frequency masking.\\ Post-processing: min time length threshold,\\predicted frames elongation.\end{tabular}\\
\hline

\end{tabular}
\caption{General characteristics of the 2021 submitted systems. Ordered by rank of F-score on the evaluation set.}
\label{tab:teams2021characteristics}
\end{table}



\clearpage

\end{document}

%% file: spectrograms/table_spectrograms.tex
\begin{figure}[]
    \centering
    \begin{tabular}{ccc}
    \includegraphics[width=0.33\textwidth]{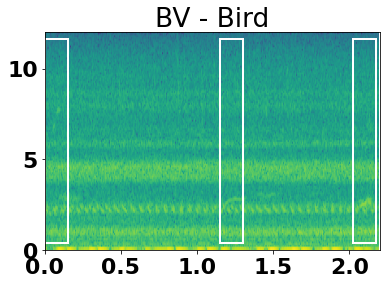}
    &
    \includegraphics[width=0.33\textwidth]{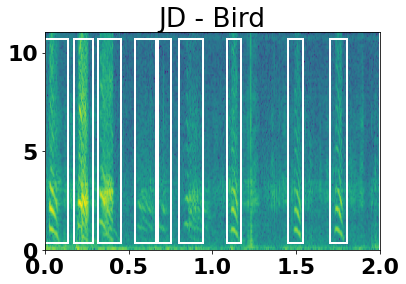}
    &
        \includegraphics[width=0.33\textwidth]{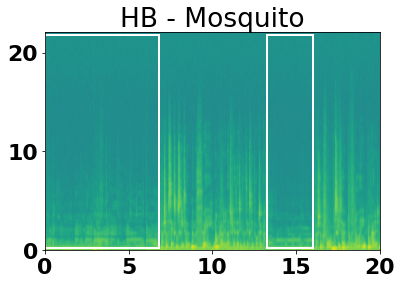}
    \\
        \includegraphics[width=0.33\textwidth]{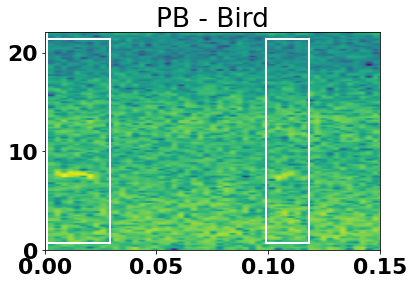}
    &
    \includegraphics[width=0.33\textwidth]{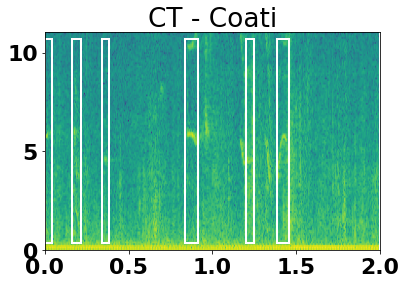}
    &
        \includegraphics[width=0.33\textwidth]{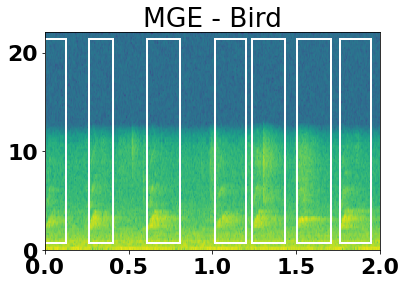}
    \\

        \includegraphics[width=0.33\textwidth]{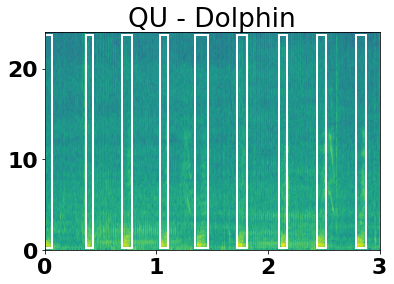}
    &
    \includegraphics[width=0.33\textwidth]{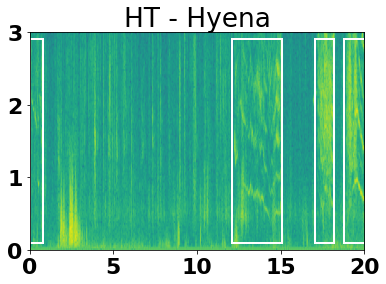}
    &

        \includegraphics[width=0.33\textwidth]{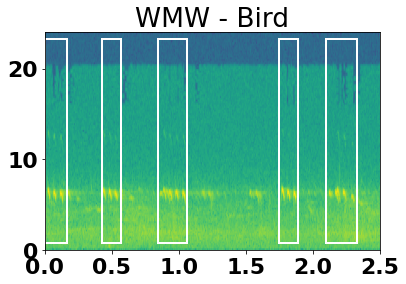}
    \\
    \includegraphics[width=0.33\textwidth]{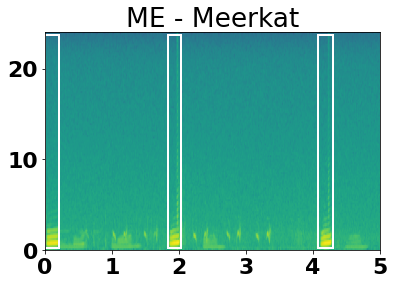}
    &
        \includegraphics[width=0.33\textwidth]{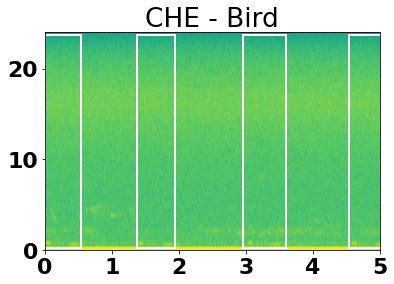}
    &
        \includegraphics[width=0.33\textwidth]{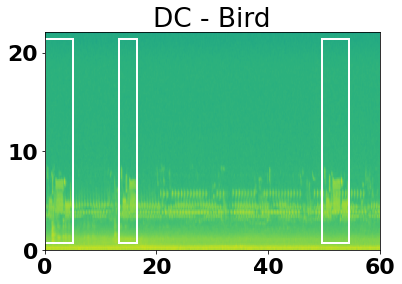}
    
    \\
        \includegraphics[width=0.33\textwidth]{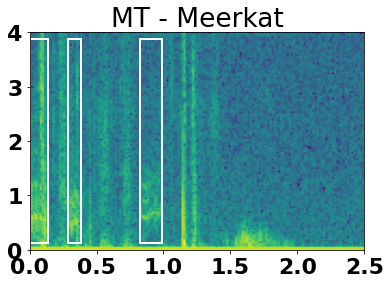}
        &
        \includegraphics[width=0.33\textwidth]{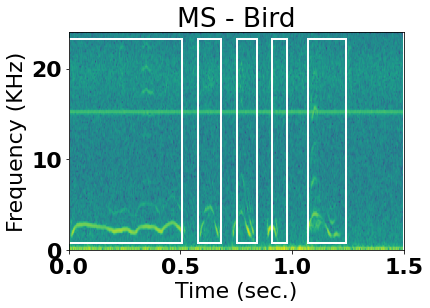}

    \end{tabular}
    \caption{Sample spectrograms for each dataset. POS (positive, i.e.\ target) vocalizations are indicated with a white rectangle.}
    \label{tab:spectrograms}
\end{figure}